\documentclass[twocolumn]{aastex61}
\usepackage{acronym,epstopdf,graphicx,hyperref}
\begin{document}

\title{Searches for Continuous Gravitational Waves from Fifteen Supernova
Remnants and Fomalhaut~b with Advanced LIGO}

\author{B.~P.~Abbott}
\affiliation{LIGO, California Institute of Technology, Pasadena, CA 91125, USA}
\author{R.~Abbott}
\affiliation{LIGO, California Institute of Technology, Pasadena, CA 91125, USA}
\author{T.~D.~Abbott}
\affiliation{Louisiana State University, Baton Rouge, LA 70803, USA}
\author{S.~Abraham}
\affiliation{Inter-University Centre for Astronomy and Astrophysics, Pune 411007, India}
\author{F.~Acernese}
\affiliation{Universit\`a di Salerno, Fisciano, I-84084 Salerno, Italy}
\affiliation{INFN, Sezione di Napoli, Complesso Universitario di Monte S.Angelo, I-80126 Napoli, Italy}
\author{K.~Ackley}
\affiliation{OzGrav, School of Physics \& Astronomy, Monash University, Clayton 3800, Victoria, Australia}
\author{C.~Adams}
\affiliation{LIGO Livingston Observatory, Livingston, LA 70754, USA}
\author{R.~X.~Adhikari}
\affiliation{LIGO, California Institute of Technology, Pasadena, CA 91125, USA}
\author{V.~B.~Adya}
\affiliation{Max Planck Institute for Gravitational Physics (Albert Einstein Institute), D-30167 Hannover, Germany}
\affiliation{Leibniz Universit\"at Hannover, D-30167 Hannover, Germany}
\author{C.~Affeldt}
\affiliation{Max Planck Institute for Gravitational Physics (Albert Einstein Institute), D-30167 Hannover, Germany}
\affiliation{Leibniz Universit\"at Hannover, D-30167 Hannover, Germany}
% ...

\collaboration{The LIGO Scientific Collaboration and the Virgo Collaboration}

\begin{abstract}
We describe directed searches for continuous gravitational waves from sixteen
well localized candidate neutron stars assuming none of the stars has a binary
companion.
The searches were directed toward fifteen supernova remnants and Fomalhaut~b,
a \relax{directly imaged} extrasolar planet candidate which has been
suggested to be a nearby old neutron star.
Each search covered a broad band of frequencies and first and second time
derivatives.
After coherently integrating spans of data from the first Advanced LIGO
observing run of 3.5--53.7~days per search, applying data-based vetoes and
discounting known instrumental artifacts, we found no astrophysical signals.
We set upper limits on intrinsic gravitational wave strain as strict as
$1\times10^{-25}$, on fiducial neutron star ellipticity as strict as
$2\times10^{-9}$, and on fiducial $r$-mode amplitude as strict as
$3\times10^{-8}$.
\end{abstract}


\acrodef{CCO}{central compact object}
\acrodef{GW}{gravitational wave}
\acrodef{O1}{the first Advanced LIGO observing run}
\acrodef{PWN}{pulsar wind nebula}
\acrodefplural{PWN}[PWNe]{pulsar wind nebulae}
\acrodef{S6}{the sixth LIGO science run}
\acrodef{SFT}{short Fourier transform}
\acrodef{SNR}{supernova remnant}

\section{Introduction}

With the detections of several binary black hole mergers \citep{GW150914,
GW151226, GW170104, GW170814, GW170608} and one binary neutron star merger
\citep{GW170817} seen also in electromagnetic waves \citep{MultiMessenger},
Advanced LIGO and Virgo have spectacularly inaugurated the field of \ac{GW}
astronomy.
While the binary neutron star merger has had far-reaching implications for our
knowledge of neutron star matter \citep{De2018, GW170817EOS}, a continuous
\ac{GW} signal could teach us even more---not just about bulk properties but
internal magnetic fields, the extent and strength of crystalline phases, and
potentially other microphysics of extreme matter \citep{Decadal,
Glampedakis2017}.

Young isolated neutron stars are promising sources of continuous \acp{GW}.
The spin-downs of young pulsars are rapid enough to include significant
continuous \ac{GW} emission, as shown by the latest \ac{GW} search for known
pulsars \citep{O1Pulsars, O1PulsarsE}.
Theoretical arguments suggest that $r$-modes (oscillations dominated by the
Coriolis force) might remain unstable and detectable in neutron stars up to a
few thousand years old \citep[and references therein]{Owen2010}.
Most young \acp{SNR} do not contain known pulsars \citep{Green2014}.
On the other hand, many of these \acp{SNR} contain small \acp{PWN}, \acp{CCO},
or other well localized non-pulsing candidate neutron stars.
Also, some of these \acp{SNR} are young enough that a neutron star could not
have been kicked far, and thus the star can be considered well localized even
if it is not seen at all.
\ac{GW} searches directed at single sky positions can significantly improve on
the sensitivities of all-sky surveys, even while needing to cover a wide band
of possible \ac{GW} frequencies and first and second time derivatives due to
lack of pulsations from the object \citep{Wette2008}.
This makes non-pulsing isolated neutron stars attractive targets for
continuous \ac{GW} searches if they are well localized.

Directed \ac{GW} searches for isolated neutron stars have been published
targeting \acp{SNR} \citep{S5CasA, S5Stochastic, S6NineSNRs, Sun2016, Zhu2016,
O1Stochastic} and promising locations including the galactic center
\citep{S5Stochastic, S5GalCen, O1Stochastic} and the core of a nearby globular
cluster---where multi-body interactions might effectively rejuvenate some
neutron stars' continuous \ac{GW} emission \citep{S6Globular}.
The only such search of data from advanced interferometers so far
\citep{O1Stochastic} employed methods from stochastic background searches
which, while quick to implement, are not as sensitive as continuous wave
search methods.

Here we present the first directed continuous wave searches for isolated
non-pulsing neutron stars in data from \ac{O1}.
We used an extension of the coherent data analysis pipeline used in
\citet{S5CasA} and \citet{S6NineSNRs}, to which this paper is a sequel.
The improved noise curve (with respect to initial LIGO and Virgo) means that
we can search more targets with sensitivity beating the indirect upper limit
on \ac{GW} emission due to energy conservation \citep{Wette2008} based on the
age of the neutron star (similar to the spin-down limit for known pulsars).
We include not only more supernova remnants, but also the \relax{directly
imaged} exoplanet candidate Fomalhaut~b, which has been proposed to be an old
nearby neutron star \citep{Fomb}---close enough that it is an attractive
target in spite of being much older than the others.
We do not include SN~1987A because it is so young that the possible spin-down
parameter space is too large to cover with a coherent wide band search and a
reasonable computational cost.

\section{Searches}

\subsection{Methods}

These searches were based on the multi-interferometer $\mathcal{F}$-statistic
\citep{Jaranowski1998, Cutler2005}.
The $\mathcal{F}$-statistic accounts for the modulation of the signal due to
the daily rotation of the detectors by adding the outputs of sinusoidal
matched filters in quadrature.
For these searches the frequency evolution of each filter, in the reference
frame of the solar system barycenter, was given by
\begin{equation}
\label{ft}
f(t) = f + \dot f (t-t_0) + \frac{1}{2} \ddot f (t-t_0)^2,
\end{equation}
where $t_0$ is the beginning of the observation and the frequency derivatives
are evaluated at that time and in a slight abuse of notation we use a simple
$f$ for $f(t_0).$
Hence these filters are designed to detect neutron stars without binary
companions whose spin-down is not too fast (requiring third or higher
frequency derivatives) or too irregular (having significant timing noise or
glitches) during the observation.
In stationary Gaussian noise, $2\mathcal{F}$ is drawn from a $\chi^2$
distribution with four degrees of freedom, which for loud signals makes the
amplitude signal-to-noise ratio roughly $\sqrt{\mathcal{F}/2}.$
If a signal is present, the $\chi^2$ is noncentral.

We used data from LIGO \ac{O1}, but none from Virgo because that
interferometer was down for upgrades during \ac{O1}.
At the frequencies to which LIGO was most sensitive (about 100-300~Hz), the
strain noise amplitude was about 3--4 times lower than in \ac{S6}
\citep{O1Instrument}.
However there were many more spectral lines due to instrumental artifacts than
in \ac{S6}, which complicated the analysis.
We used the calibration described in \citet{O1Pulsars}, which is an update of
the first \ac{O1} calibration described in \citet{O1Calibration}.
Hence, as in \citet{O1Pulsars}, our upper limits on strain are uncertain by at
least 14\%.
Like many other continuous \ac{GW} searches, ours used data in the form of
\acp{SFT} of duration 1800~s, high pass filtered and Tukey windowed to reduce
artifacts, recording only frequencies up to 2~kHz.

While each search targeted a specific direction (right ascension and
declination), each had to cover a broad band of frequencies and first and
second derivatives.
That is, a bank of signal templates was required, constructed to cover the
parameter space $(f,\dot f,\ddot f)$ with sufficient density
\citep{Whitbeck2006, Wette2008}.
We chose coverage such that the maximum loss of power signal-to-noise ratio
due to mismatch between the signal and the nearest template \citep{Owen1996,
Brady1998} was no worse than 20\%, a common choice in continuous \ac{GW}
analyses.
Given the parameter choices described below, this resulted in
$10^{12}$--$10^{13}$ templates for most searches, with the Cas~A search
getting more than $10^{14}$ since it was allocated ten times the computing
cycles of each other search.

All searches ran on the Atlas computing cluster at the Max Planck Institute
for Gravitational Physics (Albert Einstein Institute) in Hanover, Germany
using the same tag (\texttt{S6SNRSearch}) of the LALSuite software
package~\citep{lalsuite} as in \citet{S6NineSNRs} although the controlling
scripts were upgraded.
Most searches used roughly $10^5$ core hours (split into roughly $3\times10^4$
batch jobs) and Cas~A used more than $10^6$ (split into roughly $3\times10^5$
jobs).
The splitting into jobs was used in the vetoes and other post-processing
described in Sec.~\ref{s:post}.
Post-processing for each search used at most of order ten percent of the core
hours dedicated to the search.
Several terabytes of search results were written to disk.

\subsection{Target List}

\begin{table*}
\caption{
\label{t:targets}
Targeted objects and astronomical parameters used in each search
}
\begin{center}
\begin{tabular}{rcllcc}
\hline
\hline
SNR & parameter & Other name & RA+dec & $D$ & $a$ \\
(G name) & space & & (J2000) & (kpc) & (kyr) \\
\tableline
1.9$+$0.3 & & --- & 174846.9$-$271016 & 8.5 & 0.1\hspace{0.5em}
\\
15.9$+$0.2 & & --- & 181852.1$-$150214 & 8.5 & 0.54
\\
18.9$-$1.1 & & --- & 182913.1$-$125113 & 2\hspace{0.75em} & 4.4\hspace{0.5em}
\\
39.2$-$0.3 & & 3C 396 & 190404.7$+$052712 & 6.2 & 3\hspace{1.25em}
\\
65.7$+$1.2 & & DA 495 & 195217.0$+$292553 & 1.5 & 20\hspace{1.75em}
\\
93.3$+$6.9 & & DA 530 & 205214.0$+$551722 & 1.7 & 5\hspace{1.25em}
\\
111.7$-$2.1 & & Cas A & 232327.9$+$584842 & 3.3 & 0.3\hspace{0.5em}
\\
189.1$+$3.0 & wide & IC 443 & 061705.3$+$222127 & 1.5 & 3\hspace{1.25em}
\\
189.1$+$3.0 & deep & IC 443 & 061705.3$+$222127 & 1.5 & 20\hspace{1.75em}
\\
266.2$-$1.2 & wide & Vela Jr.\ & 085201.4$-$461753 & 0.2 & 0.69
\\
266.2$-$1.2 & deep & Vela Jr.\ & 085201.4$-$461753 & 0.9 & 5.1\hspace{0.5em}
\\
291.0$-$0.1 & & MSH 11$-$62 & 111148.6$-$603926 & 3.5 & 1.2\hspace{0.5em}
\\
330.2$+$1.0 & & --- & 160103.1$-$513354 & 5\hspace{0.75em} & 1\hspace{1.25em}
\\
347.3$-$0.5 & & --- & 171328.3$-$394953 & 0.9 & 1.6\hspace{0.5em}
\\
350.1$-$0.3 & & --- & 172054.5$-$372652 & 4.5 & 0.6\hspace{0.5em}
\\
353.6$-$0.7 & & --- & 173203.3$-$344518 & 3.2 & 27\hspace{1.75em}
\\
354.4$+$0.0 & wide & --- & 173127.5$-$333412 & 5\hspace{0.75em} & 0.1\hspace{0.5em}
\\
354.4$+$0.0 & deep & --- & 173127.5$-$333412 & 8\hspace{0.75em} & 0.5\hspace{0.5em}
\\
\hline
--- & wide & Fomalhaut b & 225739.1$-$293720 & \hspace{0.75em}0.011 & 316\hspace{2.25em}
\\
--- & deep & Fomalhaut b & 225739.1$-$293720 & \hspace{0.25em}0.02 & 3000\hspace{2.75em}
\\
\hline
\end{tabular}
\end{center}

Values of distance $D$ and age $a$ are generally at the optimistic (nearby and
young) end of ranges given in the literature.
For some objects the range of parameters is wide enough to justify a wide
search for optimistic parameter values (first entry for that object in the
table) and a deep search over more pessimistic parameter values (second
entry).
See text for details and references.

\end{table*}

\begin{table*}
\caption{
\label{t:targets2}
Derived parameters used in each search
}
\begin{center}
\begin{tabular}{rcrrrrrrrrr}
\hline
\hline
SNR & parameter & $f_{\min}$ & $f_{\max}$ & $T_\mathrm{span}$ &
$T_\mathrm{span}$ & Start of span & H1 & L1 & Duty & $h_0^\mathrm{age}$
\\
(G name) & space & (Hz) & (Hz) & (s) & (days) & (UTC, 2015) & SFTs & SFTs &
factor & $(\times10^{-25})$
\\
\tableline
1.9$+$0.3 & & 38 & 1332 & 336\,307 & 3.9 & Nov 30 03:53:08 & 156 & 141 & 0.79
& 8.4
\\
15.9$+$0.2 & & 72 & 538 & 887\,744 & 10.3 & Nov 25 13:39:16 & 369 & 304 & 0.68
& 3.6
\\
18.9$-$1.1 & & 45 & 987 & 1\,133\,255 & 13.1 & Nov 21 00:00:40 & 462 & 346 &
0.64 & 5.4
\\
39.2$-$0.3 & & 98 & 295 & 1\,965\,780 & 22.8 & Nov 28 00:47:19 & 641 & 647
& 0.59 & 2.1
\\
65.7$+$1.2 & & 53 & 794 & 1\,932\,067 & 22.4 & Dec 14 04:52:40 & 774 & 555 &
0.62 & 3.4
\\
93.3$+$6.9 & & 41 & 1215 & 1\,051\,764 & 12.2 & Nov 25 12:39:16 & 385 & 354 &
0.63 & 5.9
\\
111.7$-$2.1 & & 31 & 1998 & 775\,855 & 9.0 & Nov 26 20:58:03 & 317 & 294 &
0.71 & 12.6
\\
189.1$+$3.0 & wide & 37 & 1547 & 803\,419 & 9.3 & Nov 26 12:43:17 & 331 & 296
& 0.70 & 8.7
\\
189.1$+$3.0 & deep & 50 & 805 & 1\,933\,867 & 22.4 & Dec 14 04:52:40 & 775 &
555 & 0.62 & 3.4
\\
266.2$-$1.2 & wide & 19 & 1998 & 462\,616 & 5.4 & Nov 28 02:17:19 & 191 & 213
& 0.79 & 136\hspace{1.75ex}
\\
266.2$-$1.2 & deep & 32 & 1998 & 799\,819 & 9.3 & Nov 26 12:43:17 & 329 & 294
& 0.70 & 11.2
\\
291.0$-$0.1 & & 42 & 987 & 788\,409 & 9.1 & Nov 26 18:28:03 & 322 & 295 & 0.70
& 5.9
\\
330.2$+$1.0 & & 53 & 731 & 851\,744 & 9.9 & Nov 25 23:39:16 & 349 & 302 & 0.69
& 4.5
\\
347.3$-$0.5 & & 27 & 1998 & 578\,325 & 6.7 & Nov 28 05:17:19 & 237 & 253 &
0.76 & 19.9
\\
350.1$-$0.3 & & 42 & 1038 & 637\,577 & 7.4 & Nov 28 02:17:19 & 257 & 271 &
0.75 & 6.5
\\
353.6$-$0.7 & & 132 & 275 & 3\,762\,662 & 43.5 & Nov 21 02:30:40 & 1339 & 1078 &
0.58 & 1.4
\\
354.4$+$0.0 & wide & 36 & 1677 & 301\,250 & 3.5 & Nov 28 02:17:19 & 125 & 152
& 0.83 & 14.4
\\
354.4$+$0.0 & deep & 62 & 635 & 790\,209 & 9.1 & Nov 26 17:58:03 & 323 & 295 &
0.70 & 4.0
\\
\hline
Fomalhaut b & wide & 19 & 1998 & 2\,492\,267 & 28.8 & Sep 18 20:08:24 & 955 &
799 & 0.63 & 116\hspace{1.75ex}
\\
Fomalhaut b & deep & 22 & 1998 & 4\,639\,371 & 53.7 & Nov 19 23:13:10 & 1626 &
1295 & 0.57 & 20.7
\\
\hline
\end{tabular}
\end{center}

The span reported is the final one, including the possible extension to the
end of an \protect\ac{SFT} in progress at the end of the originally requested
span.
The duty factor reported is total \protect\ac{SFT} time divided by
$T_\mathrm{span}$ divided by the number of interferometers (two).
As in the previous table, for objects with two entries the first is a wide
search (optimistic parameter estimates) and the second is a deep search
(pessimistic parameter estimates).
In some cases the frequency ranges for wide and deep searches are nearly
identical, but the ranges of spin-down parameters (described in the text) are
not.

\end{table*}

\begin{figure*}
\begin{center}
\includegraphics[width=0.49\textwidth]{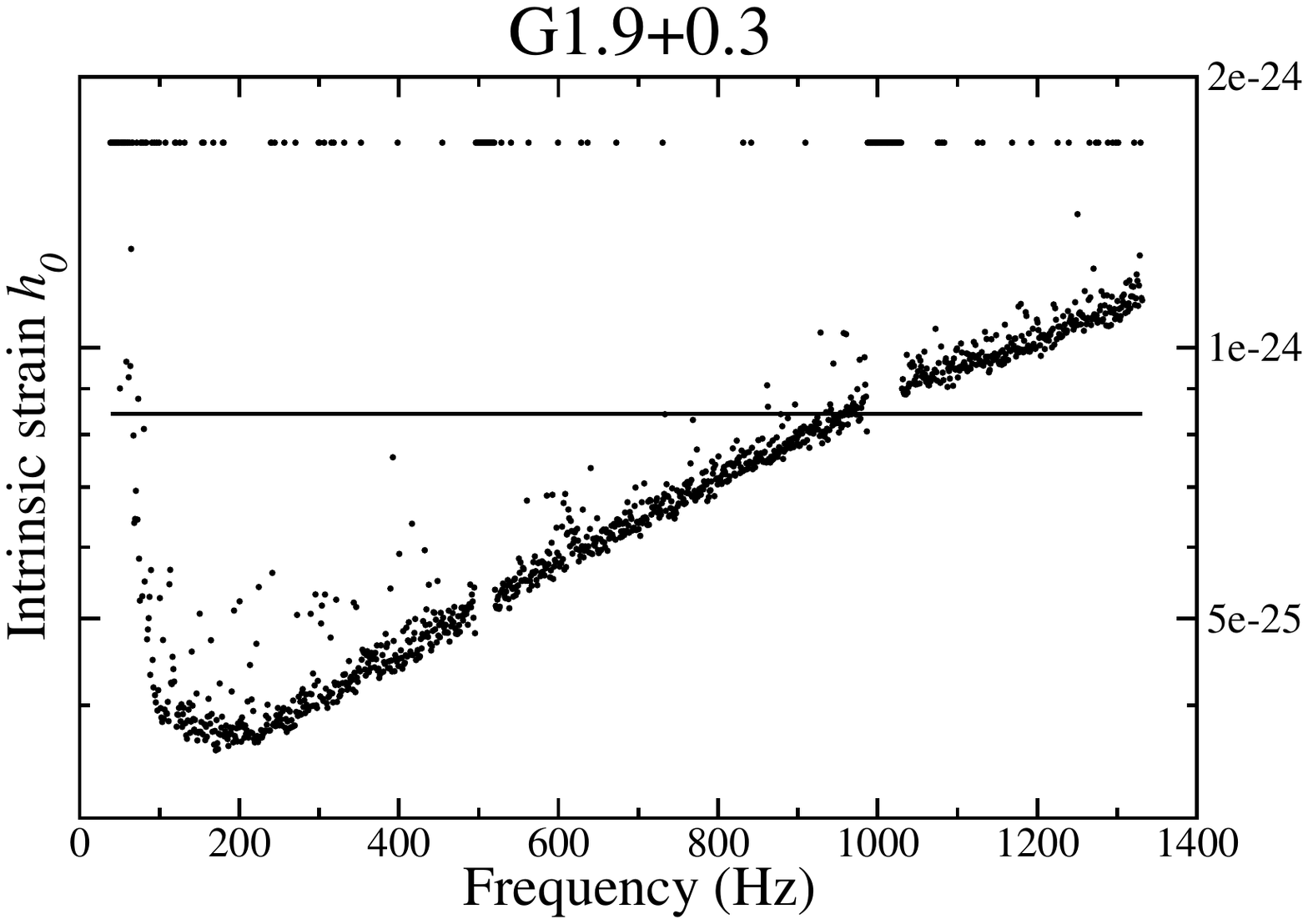}
\includegraphics[width=0.49\textwidth]{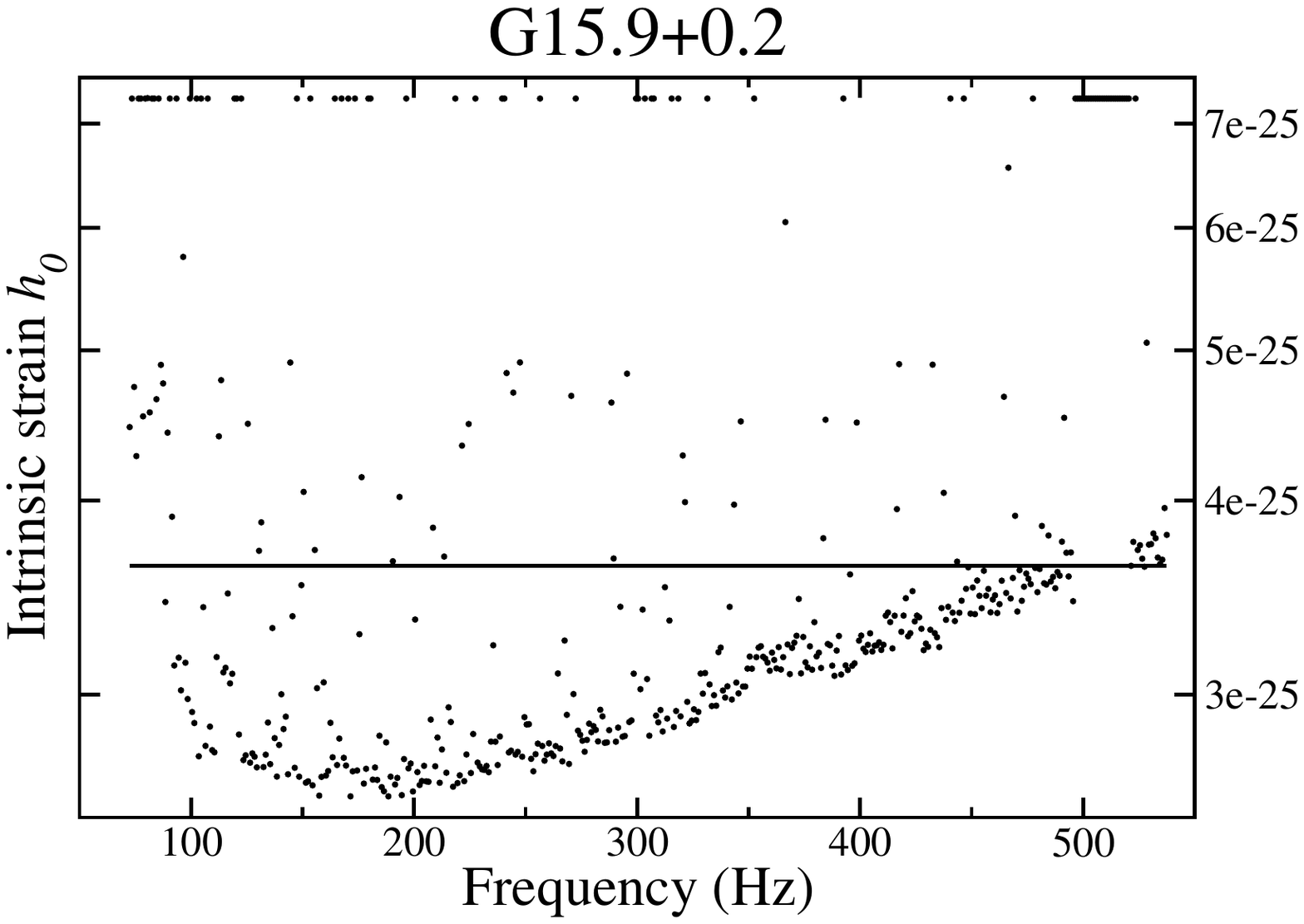}

\includegraphics[width=0.49\textwidth]{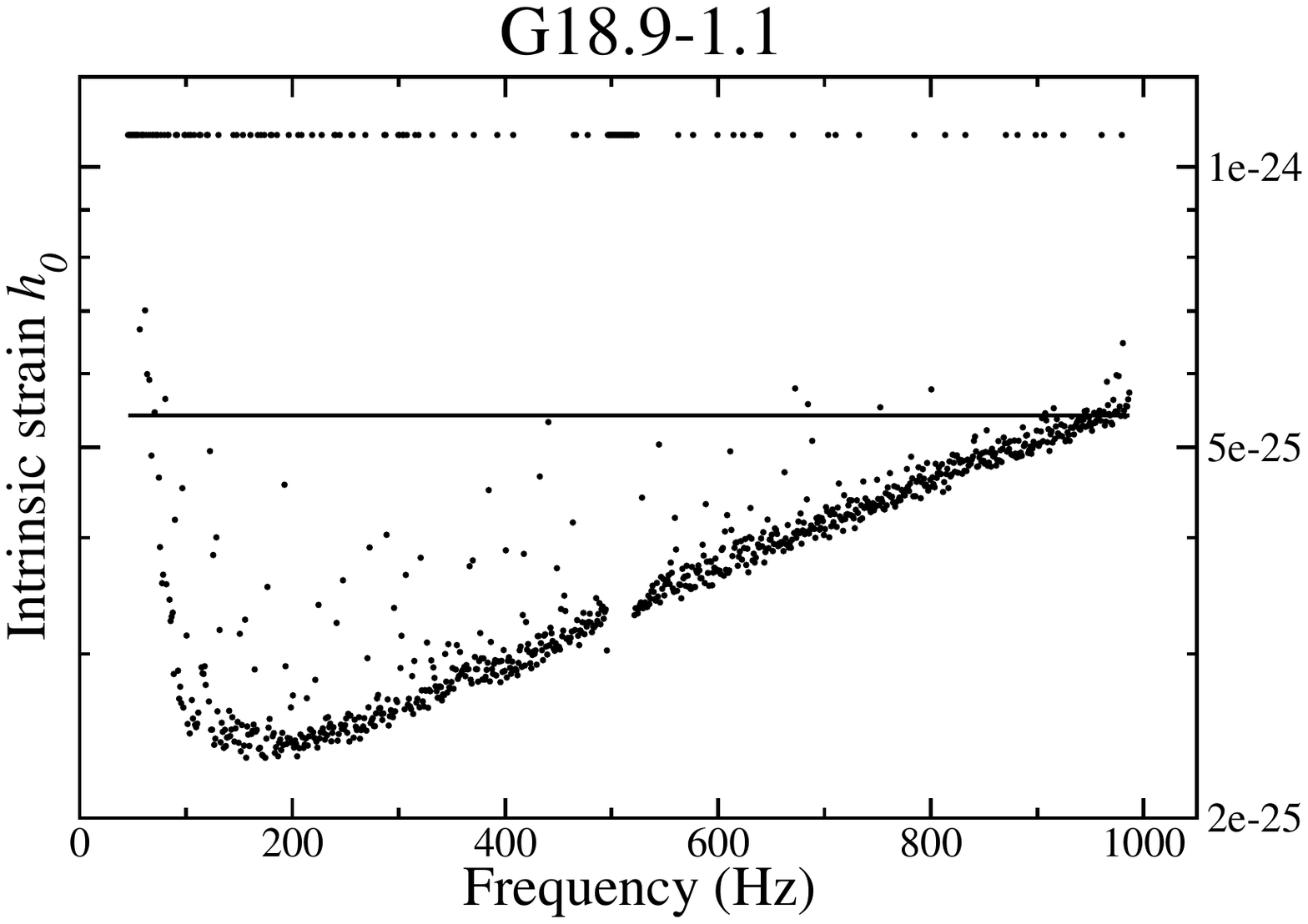}
\includegraphics[width=0.49\textwidth]{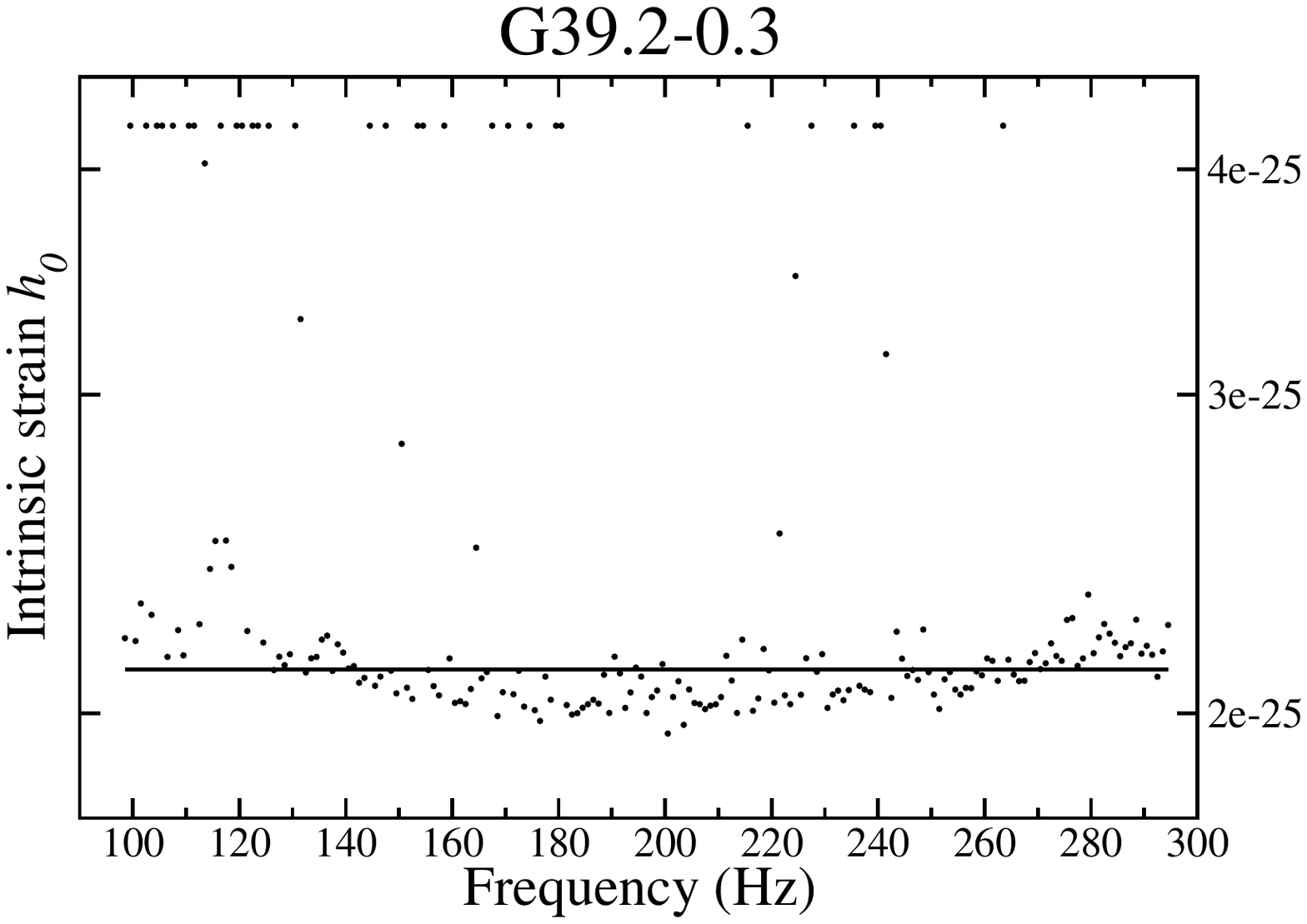}
\end{center}
\caption{
\label{f:h01}
Direct observational 95\% confidence upper limits on intrinsic strain as a
function of frequency in 1~Hz bands for four searches.
The horizontal line indicates the indirect limit from energy conservation.
Scattered points on a higher line indicate 1~Hz bands where no upper limit was
set due to data quality issues.
All figures trace a slightly distorted version of the noise curve, with
G39.2$-$0.3 appearing flat because it covers only the bottom of the curve.
}
\end{figure*}

\begin{figure*}
\begin{center}
\includegraphics[width=0.49\textwidth]{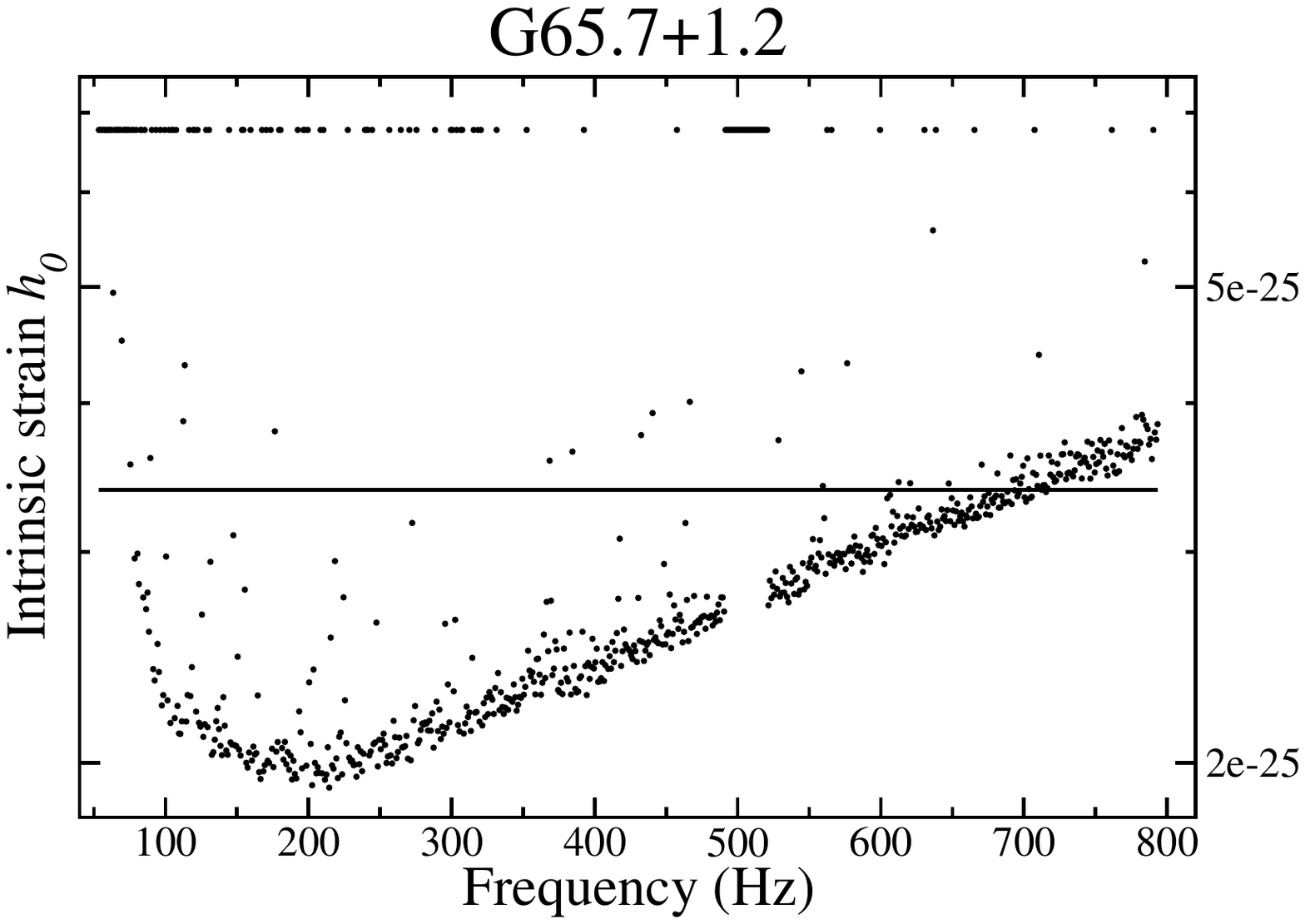}
\includegraphics[width=0.49\textwidth]{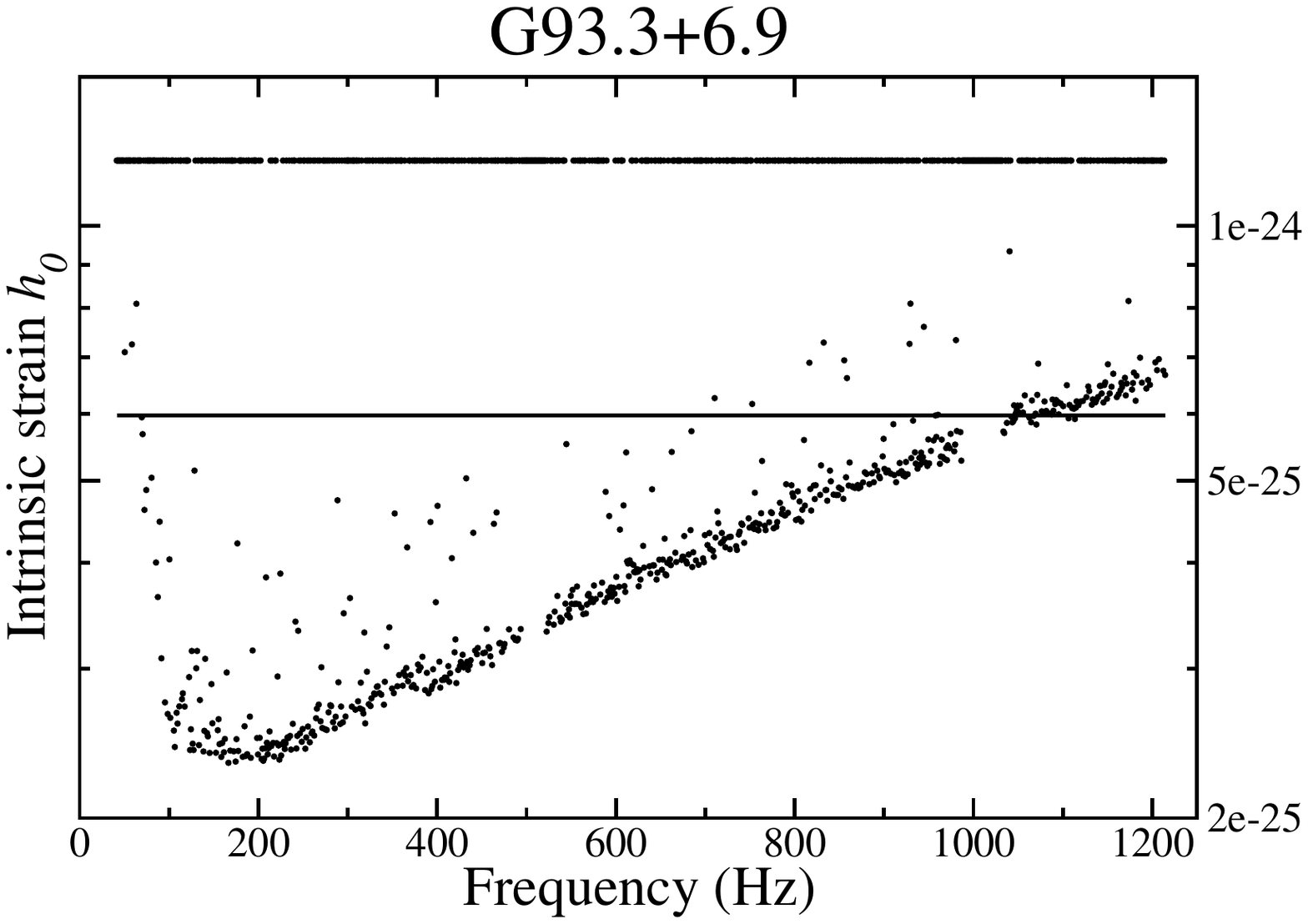}

\includegraphics[width=0.48\textwidth]{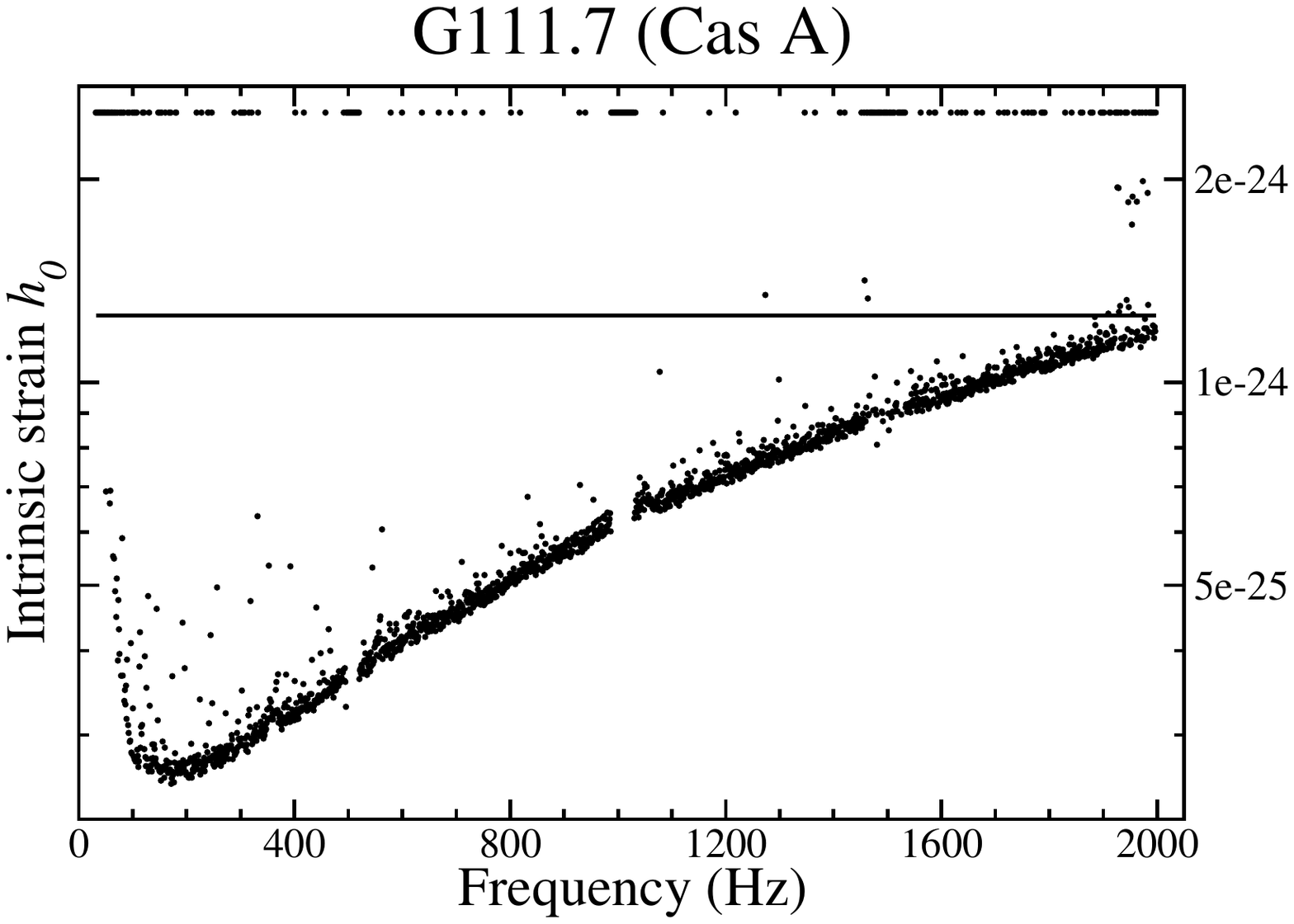}
\includegraphics[width=0.48\textwidth]{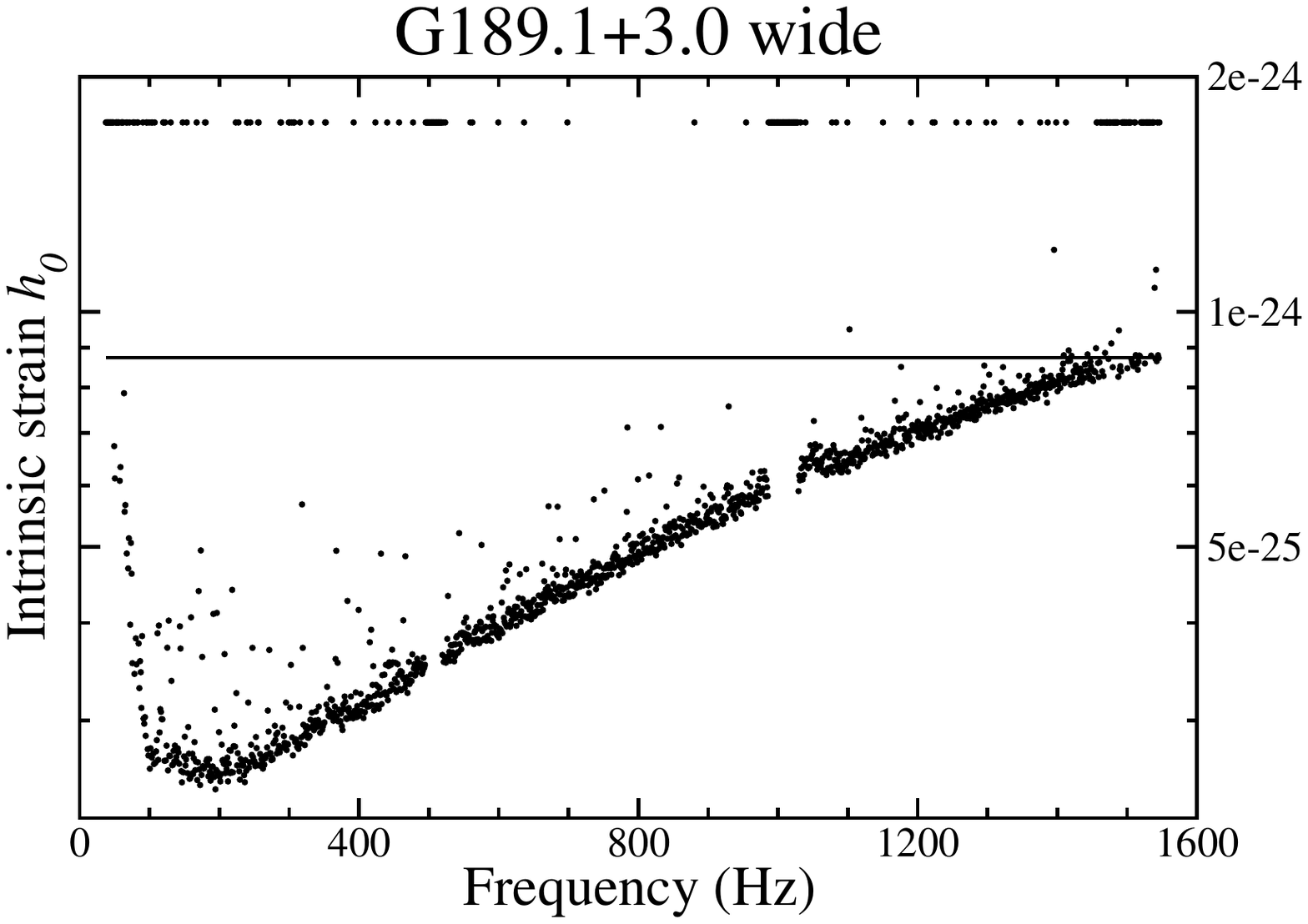}
\end{center}
\caption{
Same as the previous figure for four more searches.
}
\end{figure*}

\begin{figure*}
\begin{center}
\includegraphics[width=0.49\textwidth]{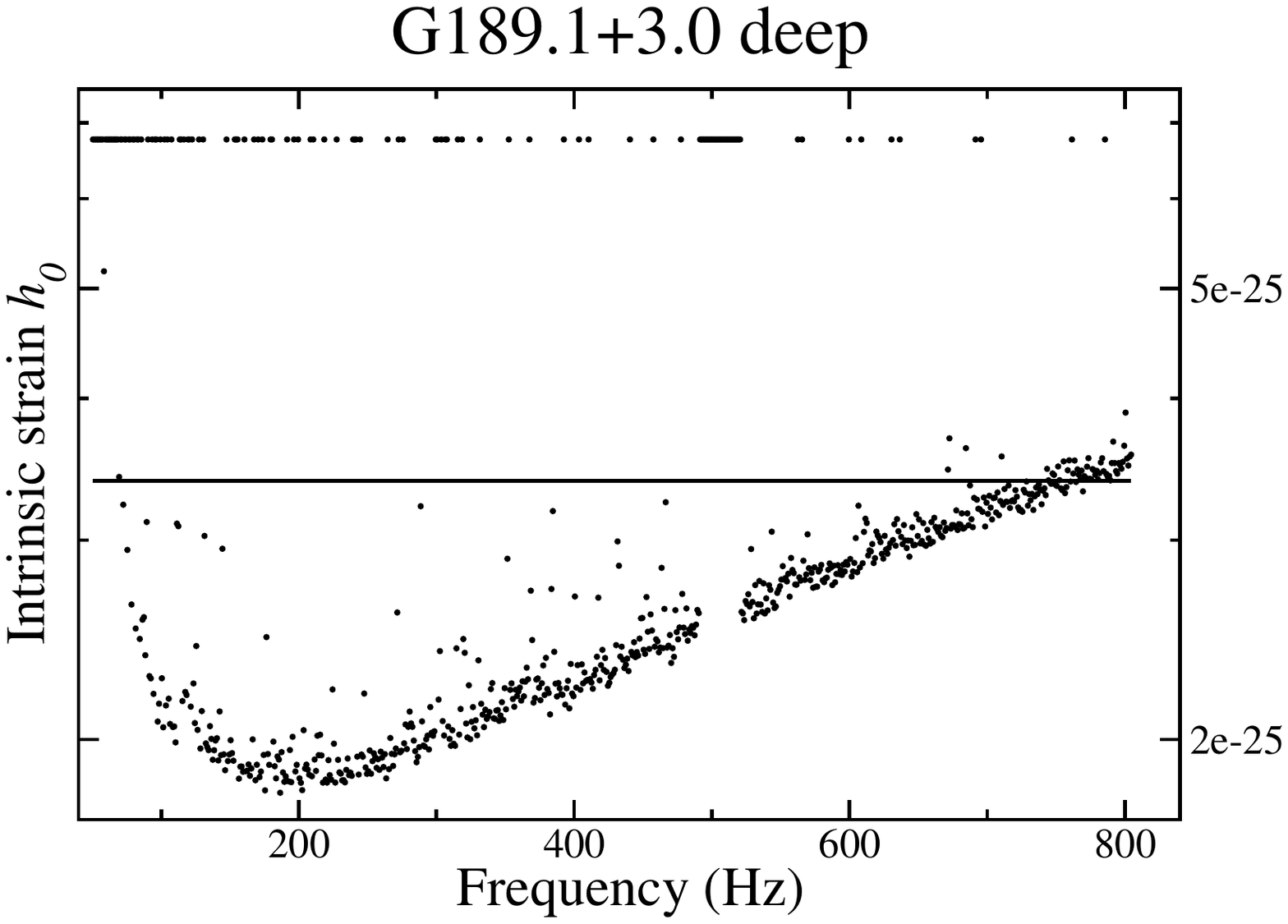}
\includegraphics[width=0.49\textwidth]{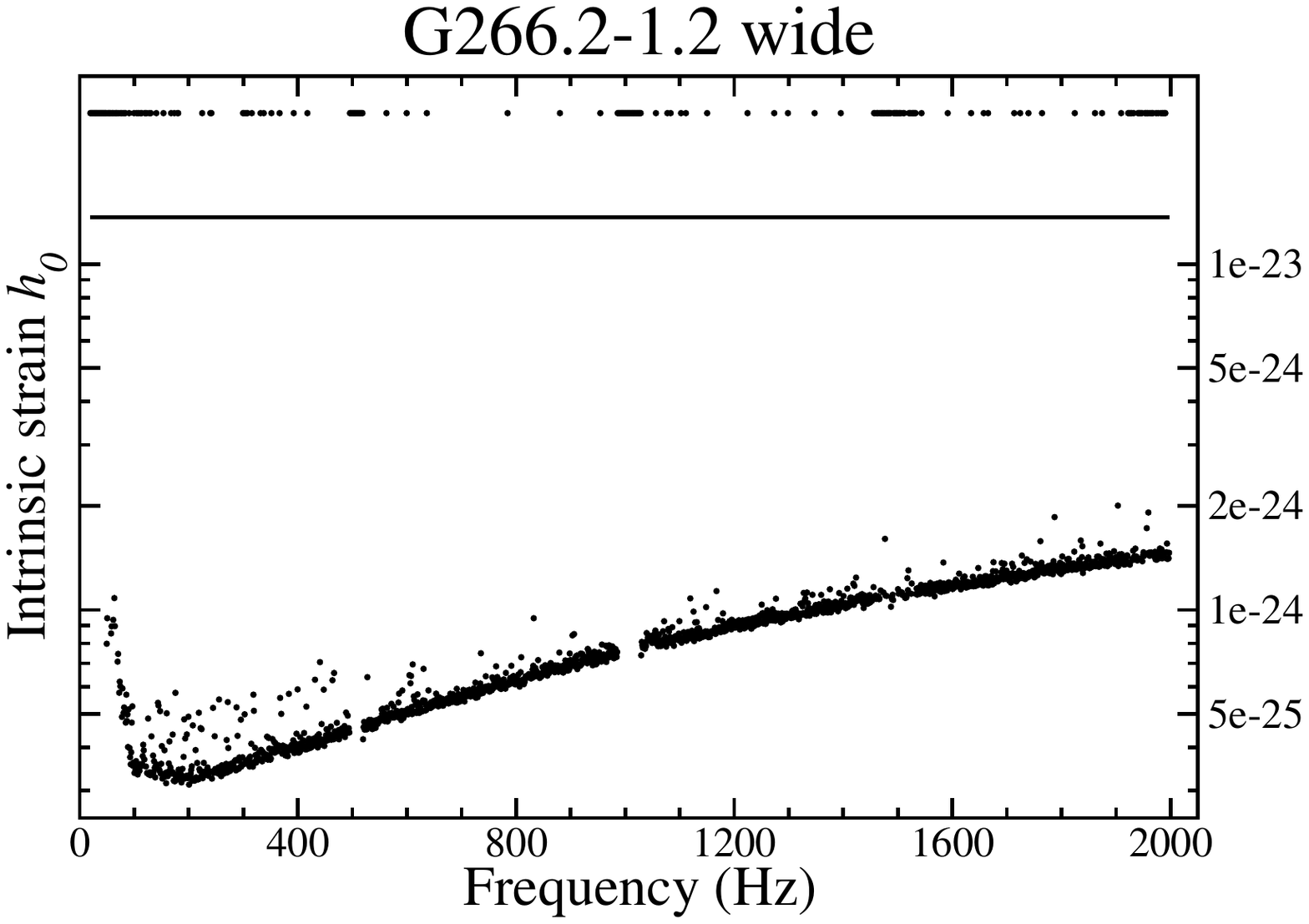}

\includegraphics[width=0.49\textwidth]{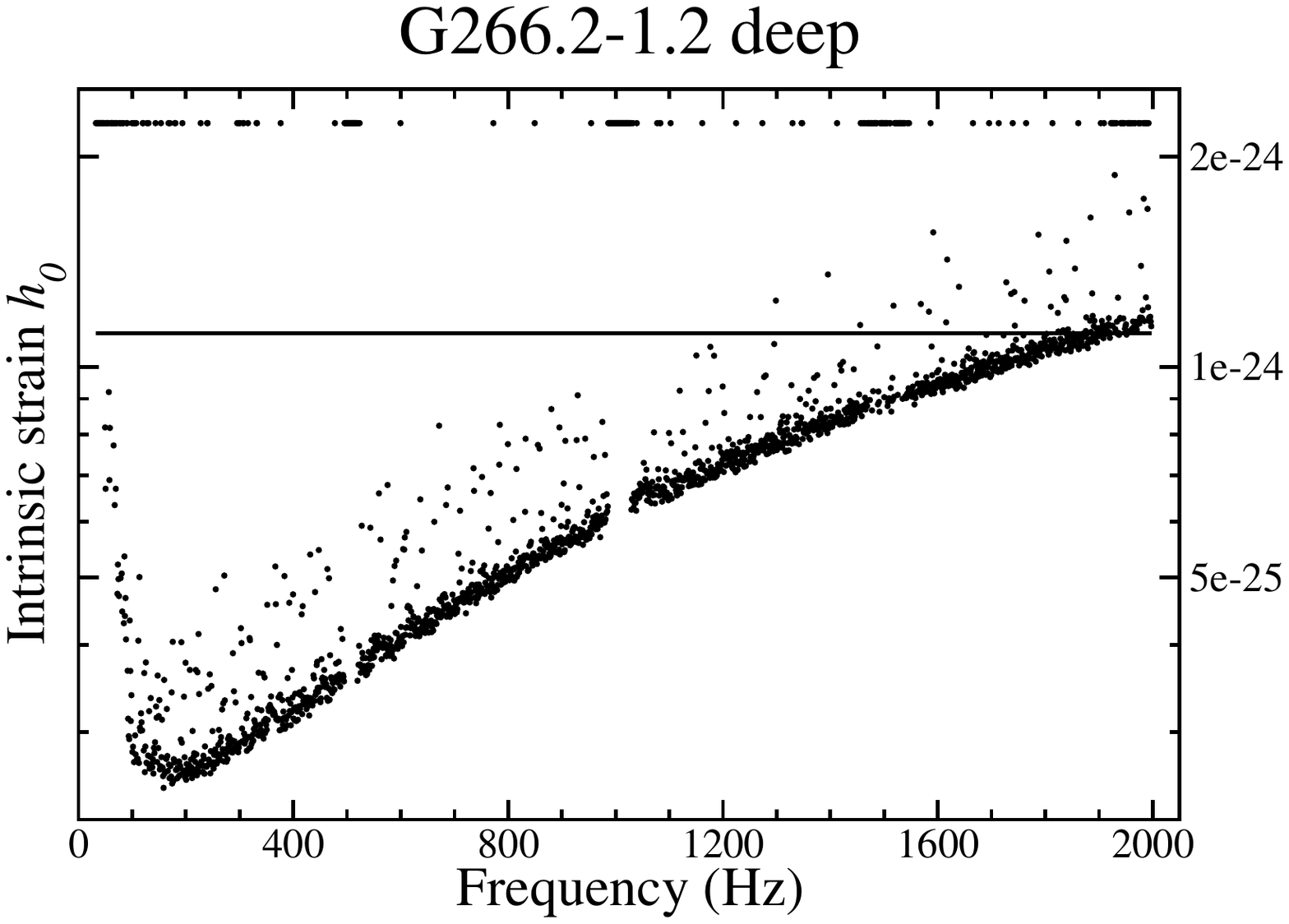}
\includegraphics[width=0.49\textwidth]{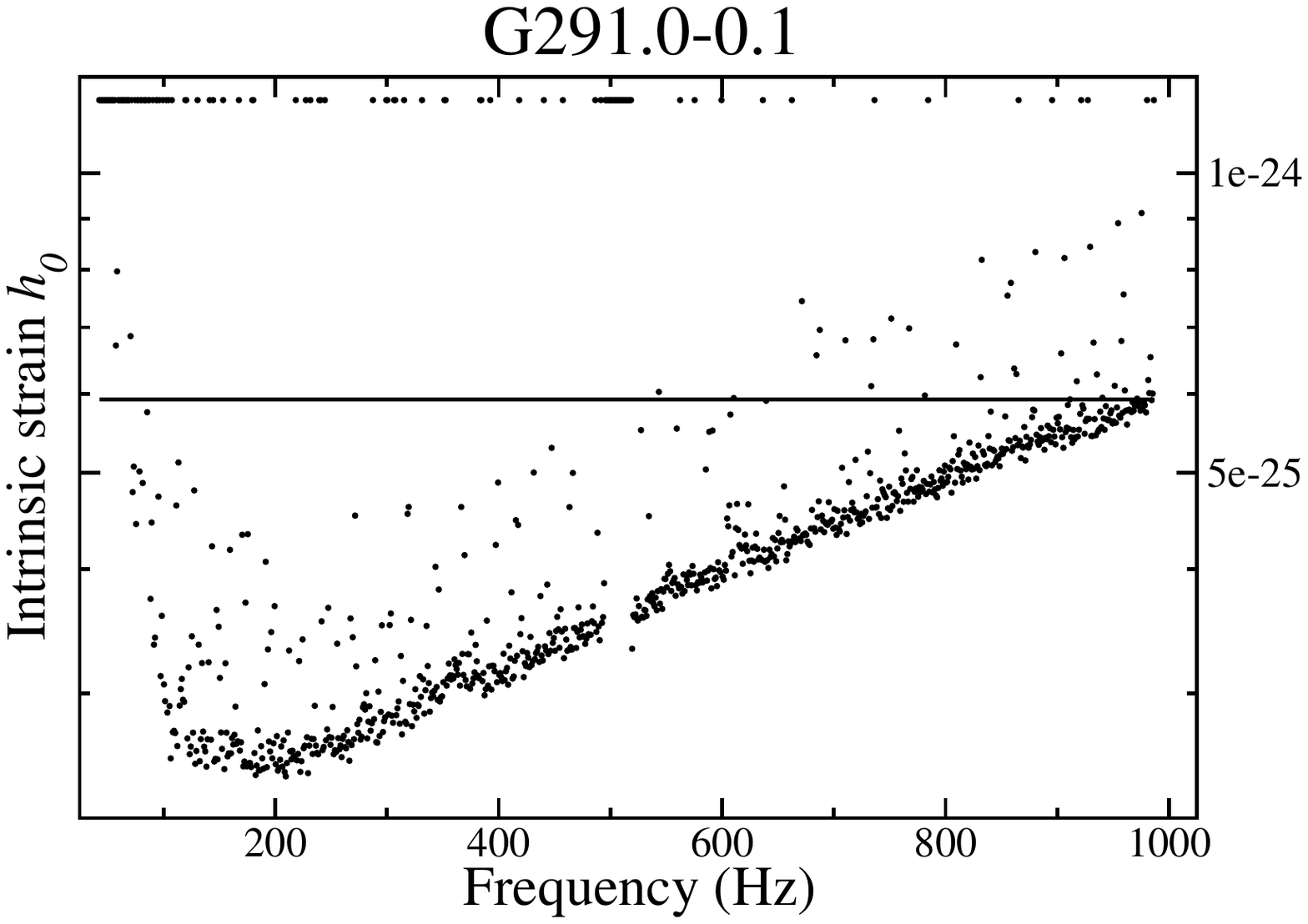}
\end{center}
\caption{
Same as the previous figure for four more searches.
}
\end{figure*}

\begin{figure*}
\begin{center}
\includegraphics[width=0.49\textwidth]{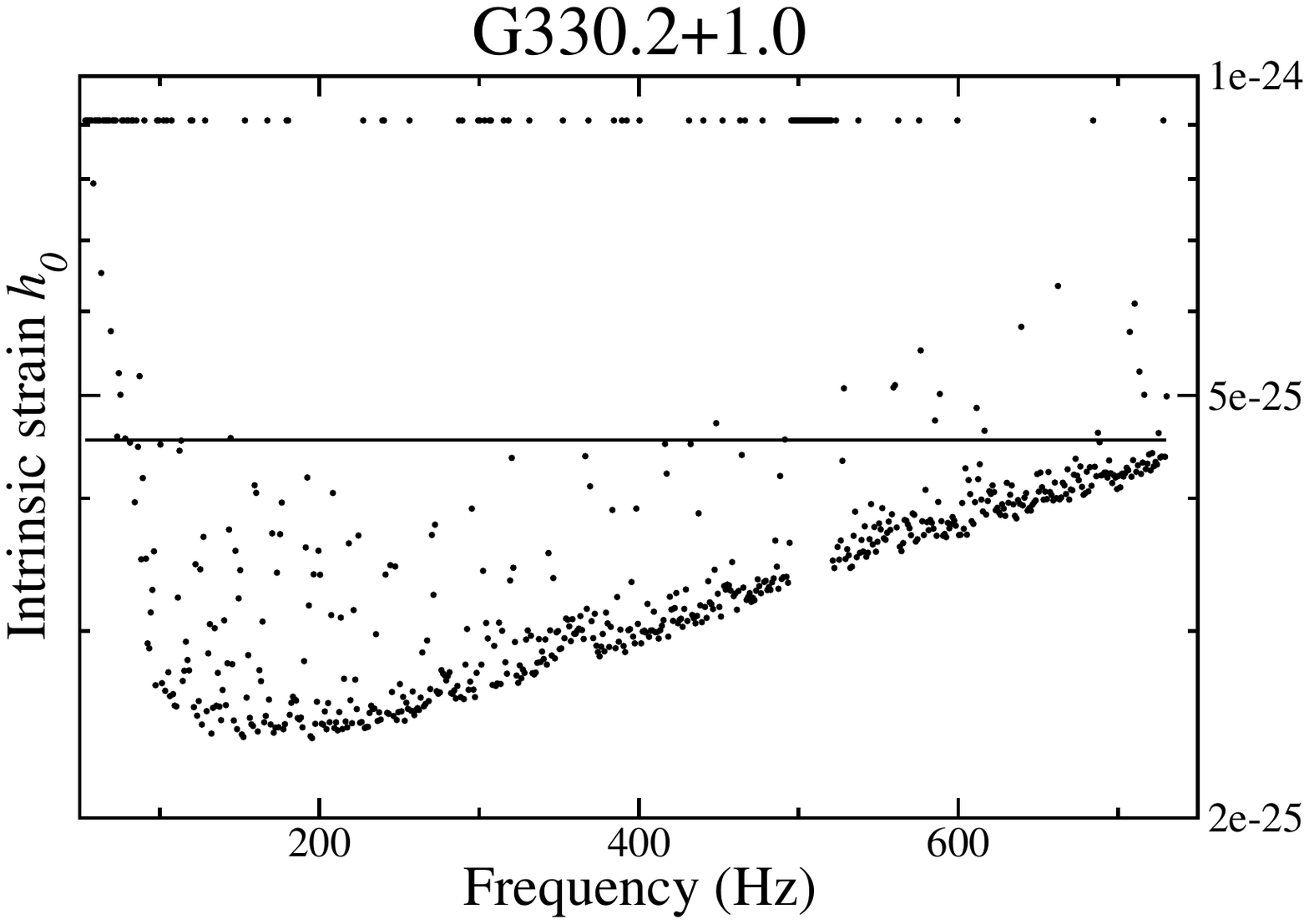}
\includegraphics[width=0.49\textwidth]{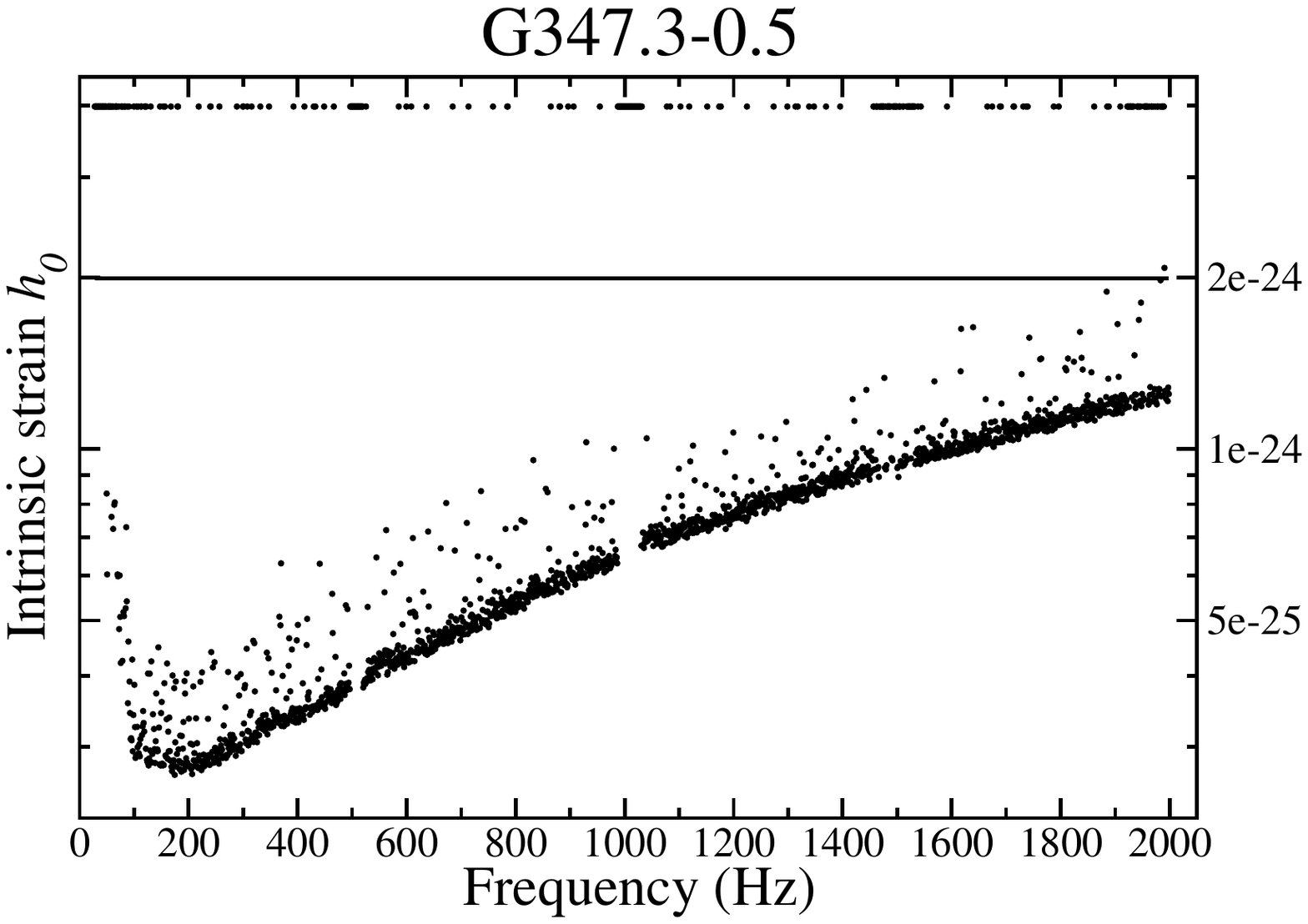}

\includegraphics[width=0.49\textwidth]{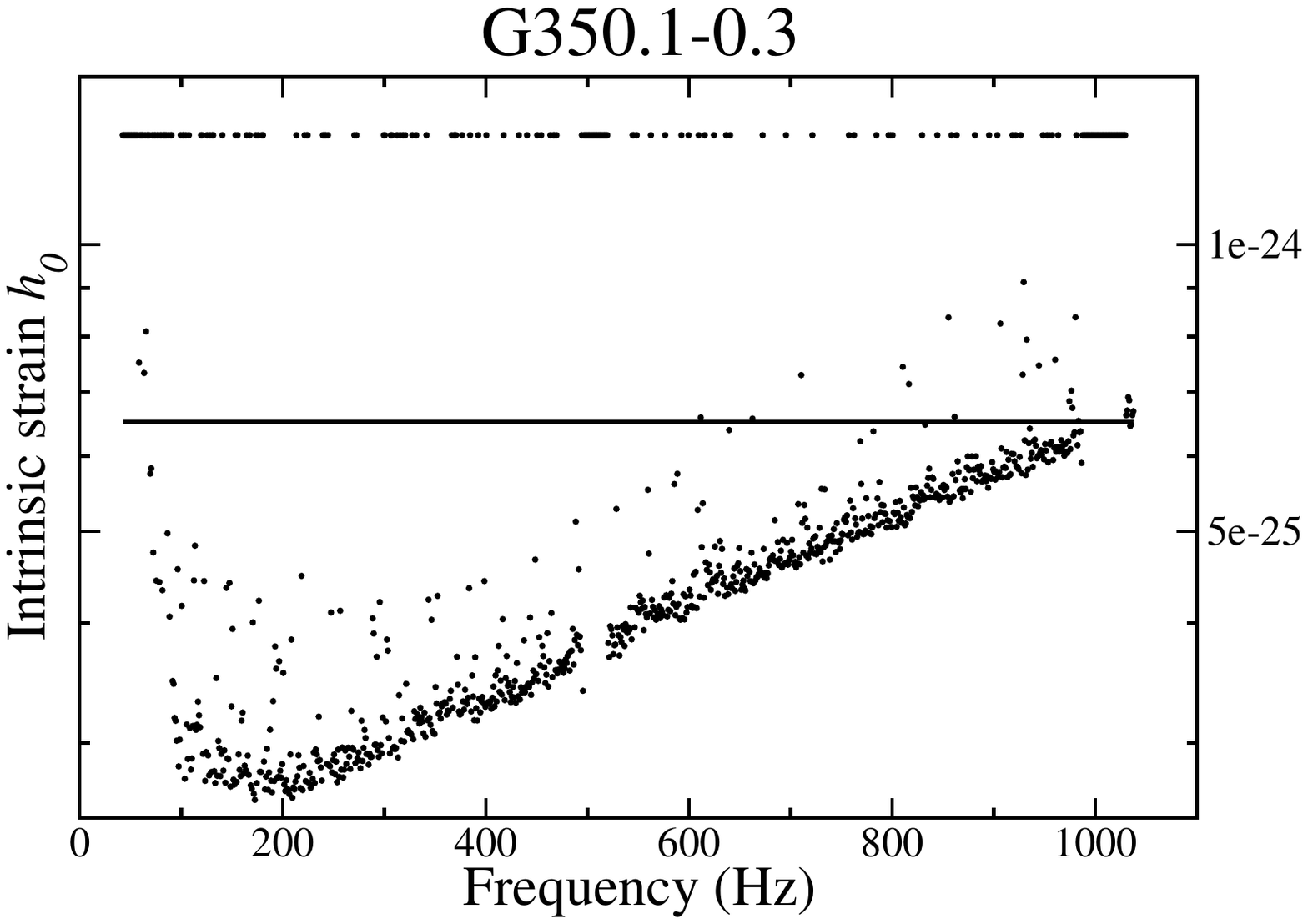}
\includegraphics[width=0.49\textwidth]{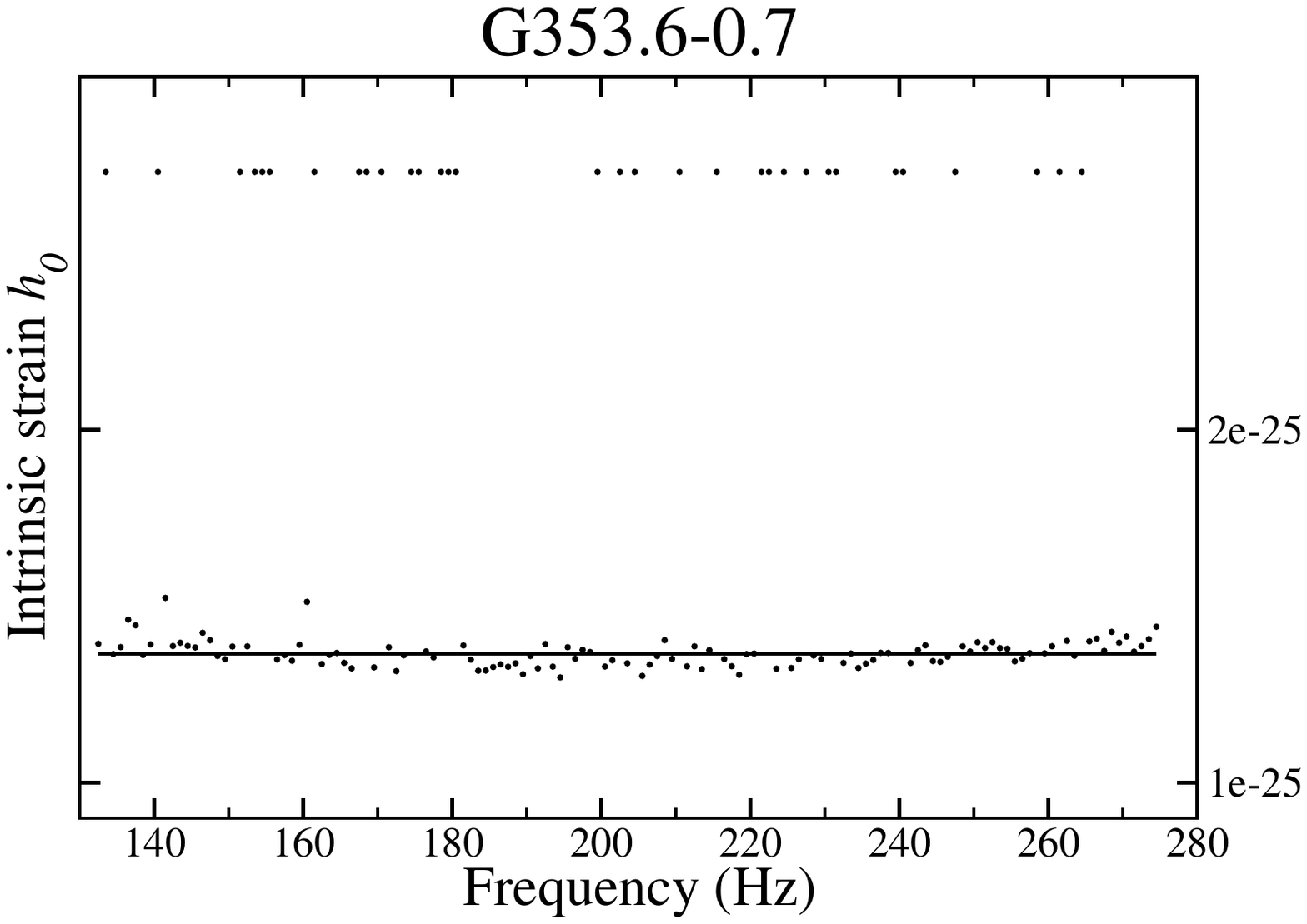}
\end{center}
\caption{
Same as the previous figure for four more searches.
}
\end{figure*}

\begin{figure*}
\begin{center}
\includegraphics[width=0.49\textwidth]{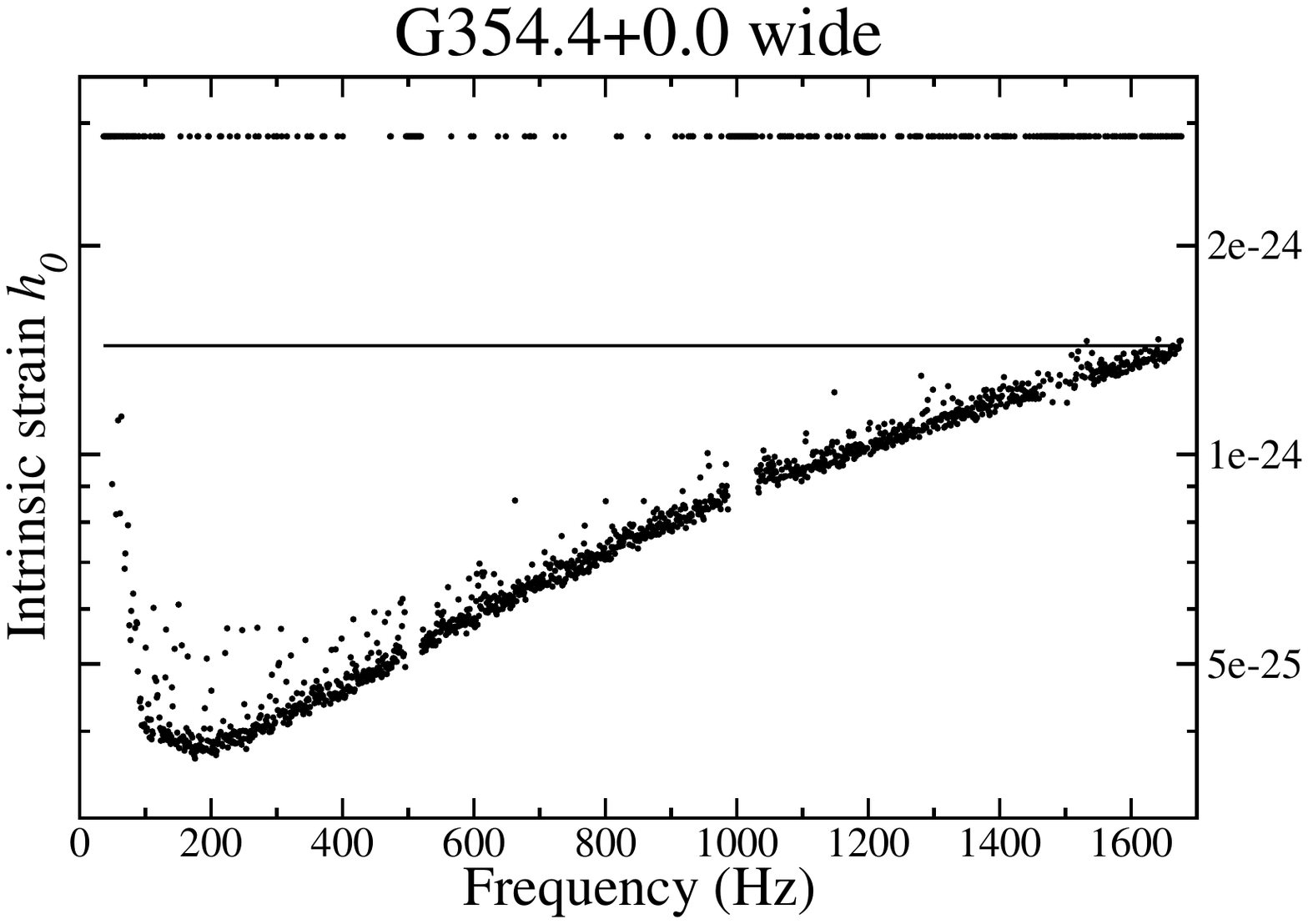}
\includegraphics[width=0.49\textwidth]{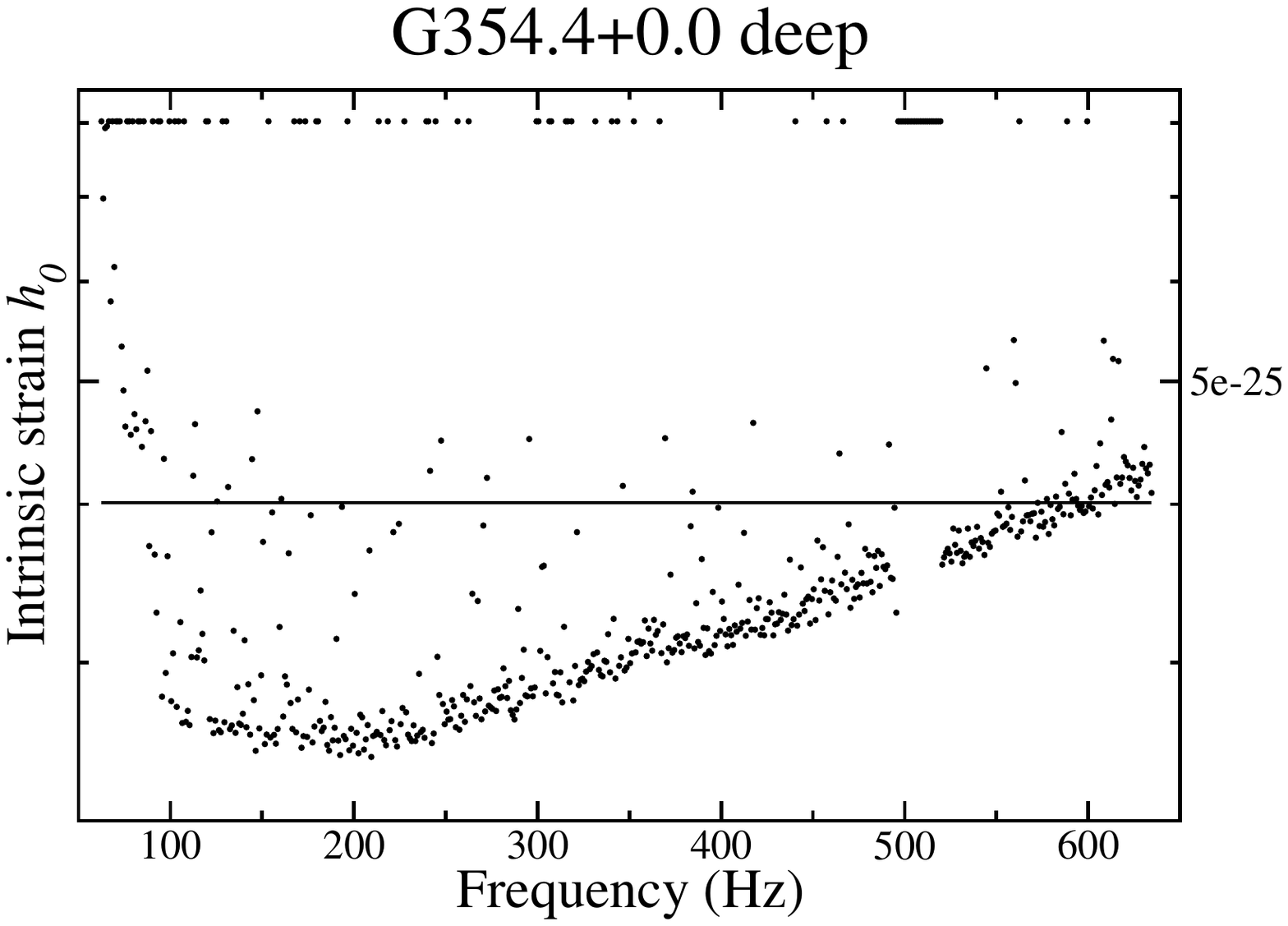}

\includegraphics[width=0.49\textwidth]{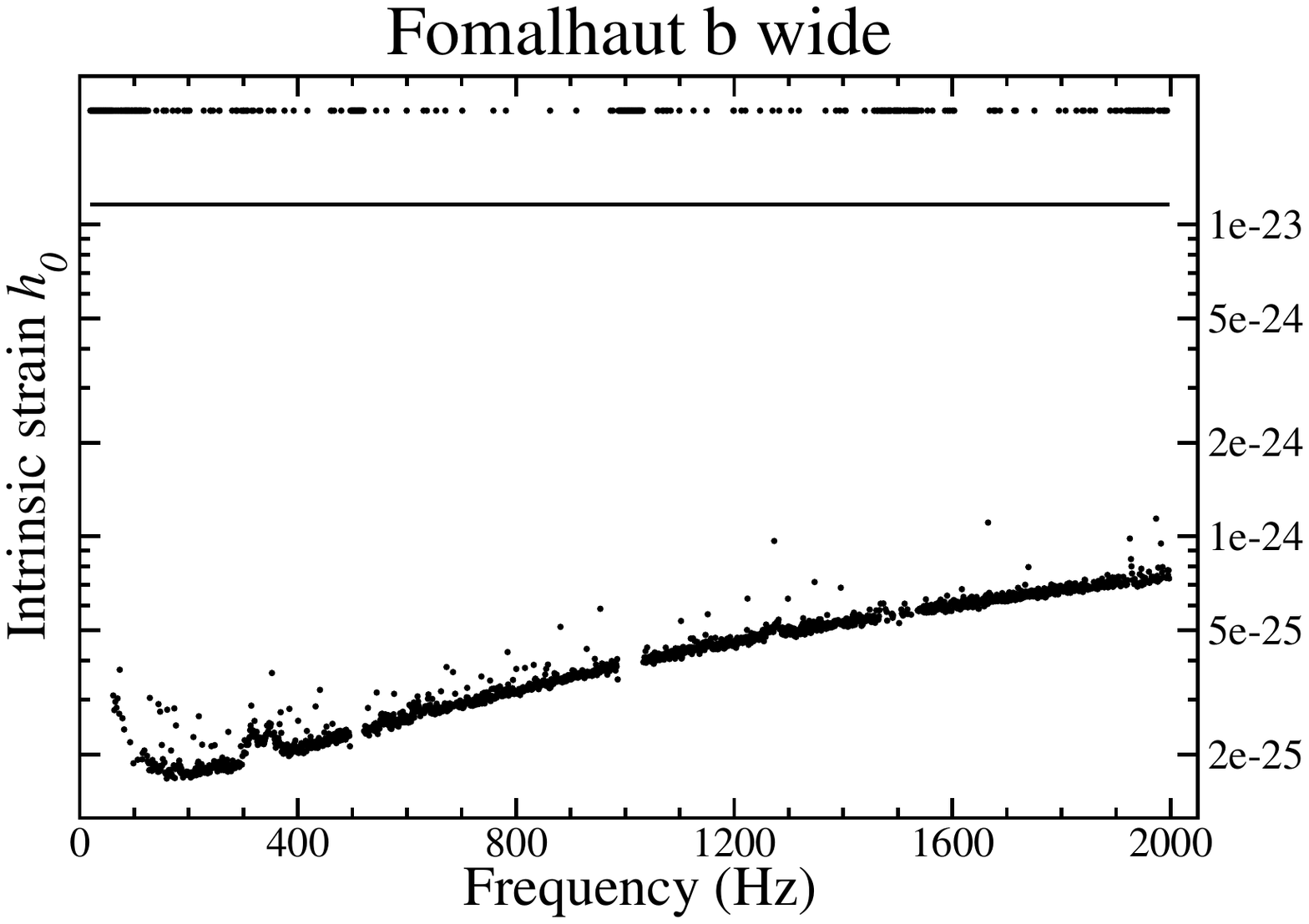}
\includegraphics[width=0.49\textwidth]{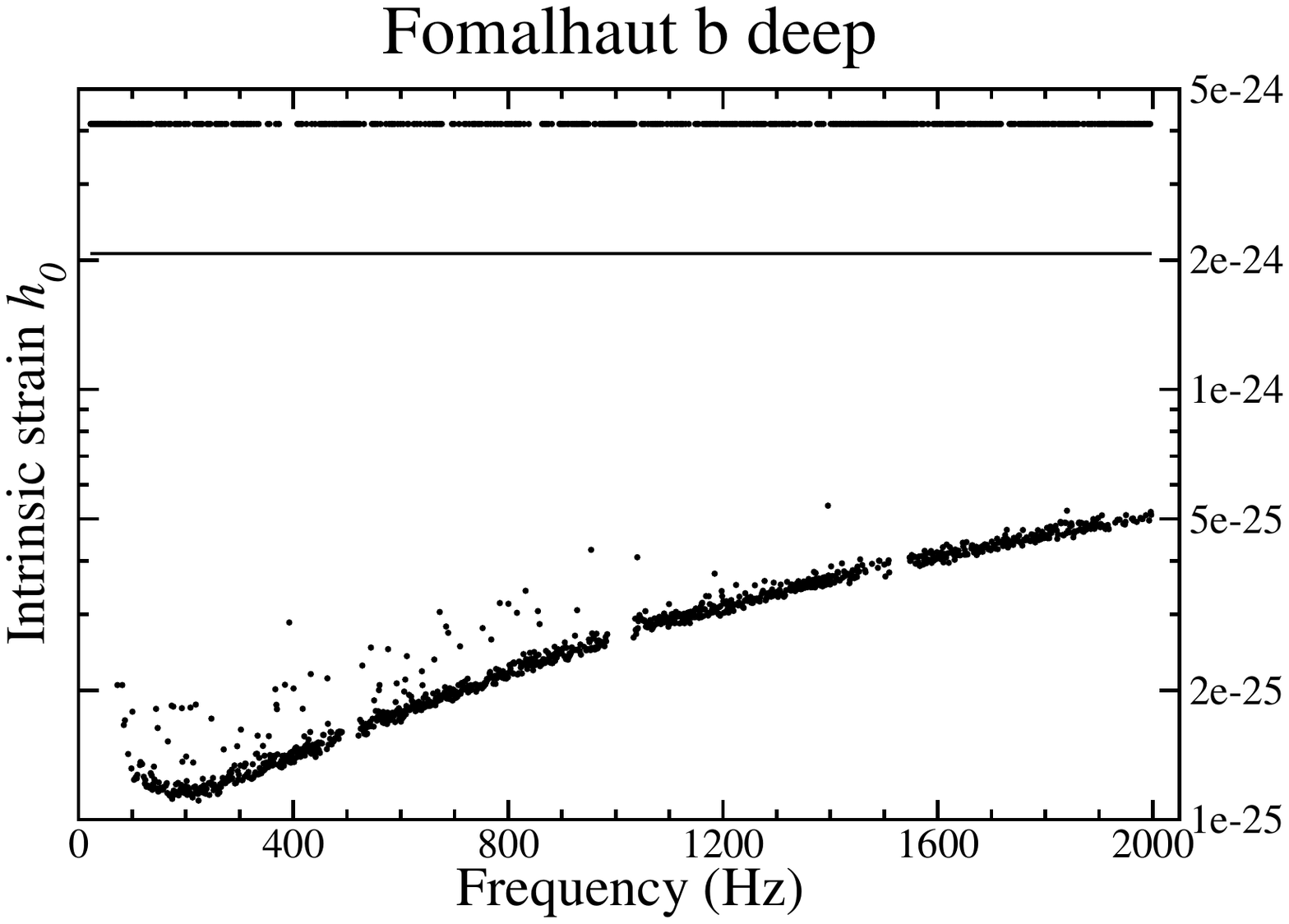}
\end{center}
\caption{
\label{f:h00}
Same as the previous figure for four more searches.
}
\end{figure*}

Our choice of targets required that a search of fixed computational cost be
sensitive enough to detect the strongest continuous \ac{GW} signal consistent
with broad conservation of energy considerations.
As introduced by \citet{Wette2008} \relax{for the \ac{CCO} in supernova
remnant Cas~A,} the strongest possible signal based on the age $a$ and
distance $D$ of the source,
\begin{equation}
h_0^\mathrm{age} = 1.26\times10^{-24} \left( \frac{\mbox{3.30 kpc}} {D}
\right) \left( \frac{\mbox{300 yr}} {a} \right)^{1/2},
\end{equation}
is analogous to the spin-down limit for known pulsars and indicates the
strongest possible intrinsic strain produced by an object whose unknown
spin-down is entirely due to \ac{GW} emission and has been since birth.
\relax{The fiducial parameters (for Cas~A) show that this limit can be high
enough to be interesting.}
The intrinsic strain $h_0$ \citep{Jaranowski1998} characterizes the \ac{GW}
metric perturbation without reference to any particular orientation or
polarization, and hence is typically a factor 2--3 greater than the strain
response measured by the interferometers.
The indirect limit $h_0^\mathrm{age}$ is slightly different for $r$-mode
emission \citep{Owen2010} than for the mass quadrupole source tacitly assumed
above and in most of the literature, but we neglect this small difference.
Due to uncertainties in the neutron star mass and equation of state,
$h_0^\mathrm{age}$ is uncertain by of order 50\%, which we also neglect.

To choose directions to search, we started from the Green catalog of supernova
remnants \citep{Green2014}.
We picked x-ray point sources (\acp{CCO} or candidate \acp{CCO}), small
\acp{PWN}, and in some cases relatively young \acp{SNR} where any neutron star
could not yet have moved far.
We selected only targets with age and distance estimates so that we could
evaluate $h_0^\mathrm{age}.$
In some cases there is a wide range of estimates in the literature, leading to
significant differences in $h_0^\mathrm{age}.$
In most cases we used the most optimistic estimates, yielding the highest
$h_0^\mathrm{age}$ but also the most difficult search over the widest band of
frequency and spin-down parameters.
In addition to this wide search using the optimistic age and distance, we did
a deep search using the most pessimistic age and distance in cases where the
strain sensitivity would improve over the wide search by a factor of roughly
$\sqrt{2}.$

The resulting targets and chosen parameters are shown in
Table~\ref{t:targets}.
We now briefly summarize each target and the provenance of the parameters used
for it.

\textit{G1.9+0.3}---Currently the youngest known \ac{SNR} in the galaxy
\citep{Reynolds2008}.
Several arguments favor it being a Type~Ia \citep{Reynolds2008}, which would
leave no neutron star behind, but this is not definite and the remnant's youth
makes it an interesting target on the chance that it is not Type~Ia.
We used the position of the center of the remnant from the discovery paper
\citep{Reich1984}.
At maximum kick velocity any neutron star could have moved only a few
arcseconds, which is not an issue for our searches.
The age and distance shown are from the ``rediscovery'' paper
\citep{Reynolds2008}, though the latter is a nominal galactic center distance.

\textit{G15.9+0.2}---The \ac{CCO} was discovered in \textit{Chandra} data by
\citet{Reynolds2006}.
We used the lower limit on age and the galactic center distance estimate from
the same paper, though both quantities may be significantly greater
\citep{Klochkov2016}.

\textit{G18.9--1.1}---The position is that of the \textit{Chandra} point
source discovered by \citet{Tullmann2010}.
Age and distance estimates are from the previous \textit{ROSAT} and
\textit{ASCA} observations of \citet{Harrus2004}.

\textit{G39.2--0.3}---Also known as 3C~396.
The \ac{PWN} and embedded point source were found by \citet{Olbert2003} in
\textit{Chandra} data, the point source being localized to within $2''$ in
spite of the \ac{PWN}.
\citet{Su2011} estimate the age and distance, the latter based on the tangent
point of the spiral arm.

\textit{G65.7+1.2}---Also known as DA~495.
\citet{Arzoumanian2008} found the \textit{Chandra} point source in the
\ac{PWN}.
The quoted distance \citep{Kothes2004} and minimum age \citep{Kothes2008} are
derived slightly inconsistently due to assumed distances to the galactic
center.
We did not attempt to resolve the inconsistency, though we did choose the
distance from the former paper since it uses the more commonly accepted
galactic center distance.
The latter paper (and others) also argue that the distance could be several
times higher.

\textit{G93.3+6.9}---Also known as DA~530.
The position and age are from \citet{Jiang2007} and the distance estimate is
from \citet{Foster2003}.
\citet{Jiang2007} find no \textit{Chandra} point source, but the X-ray
intensity of the faint candidate \ac{PWN} falls off on a scale of $6'',$ which
qualifies as a point source for our purposes.

\textit{G111.7--2.1}---Also known as Cas~A.
The position of the \ac{CCO} is from the \textit{Chandra} ``first light''
observation \citep{Tananbaum1999}, the distance is from \citet{Reed1995}, and
the age is from \citet{Fesen2006}.

\textit{G189.1+3.0}---Also known as IC~443.
The position is that of the \textit{Chandra} point source found by
\citet{Olbert2001} embedded in the \ac{PWN}.
This object is often studied, with a wide range of distance and age estimates
in the literature.
We used \citet{Petre1988} for an optimistic age estimate.
Our pessimistic age estimate is not quite the most extreme in the literature,
but rather a best fit for a pessimistic scenario from relatively recent
modeling \citep{Swartz2015}.
We did not use the most optimistic distance quoted, but rather the assumed
association with the I~Gem cluster from \citet{Fesen1980} which is common in
the literature.

\textit{G266.2--1.2}---Also known as Vela~Jr.
The position is that of the \ac{CCO} found by \citet{Pavlov2001}.
We used \citet{Iyudin1998} for the most optimistic age and distance estimates.
The pessimistic age estimate is from \citet{Allen2015}, which was published
too recently for the previous paper in this series \citep{S6NineSNRs}.
\citet{Allen2015} also discuss the possible association of several surrounding
objects with the nearer concentration of the Vela Molecular Ridge, at a spread
of distances providing our pessimistic distance estimate \citep{Liseau1992}
and rendering the more pessimistic ones unlikely.

\textit{G291.0--0.1}---Also known as MSH~11$-$62.
The position and age are from the \textit{Chandra} point source discovery
paper \citep{Slane2012}.
The distance is from \citet{Moffett2001}.
The age and distance are derived in slightly inconsistent ways, but rather
than attempt to repeat the calculations we used the numbers quoted in the
literature.

\textit{G330.2+1.0}---The \ac{CCO} was discovered by \citet{Park2006} in
\textit{Chandra} data with sub-arcsecond position accuracy.
We used a distance estimate from radio observations
\citep{McClure-Griffiths2001} and an age estimate from the x-ray spectrum
\citep{Park2009}.

\textit{G347.3--0.5}---\citet{Mignani2008} obtained the sub-arcsecond
position from archival \textit{Chandra} data, although the \ac{CCO} had been
identified in \textit{ASCA} data earlier \citep{Slane1999}.
We used the distance from \citet{Cassam-Chenai2004} and the age from the
proposed identification with a possible SN~393 \citep{Wang1997}.
Although this identification may be problematic given the inferred properties
of such a supernova, other age estimates are comparable \citep{Fesen2012}.

\textit{G350.1--0.3}---Position and distance estimates are from the
discovery paper of the \ac{CCO} candidate by \citet{Gaensler2008}.
The age is from \textit{Chandra} observations \citet{Lovchinsky2011}.

\textit{G353.6--0.7}---\citet{Halpern2010} identified the most likely of
several candidate \acp{CCO}.
The age estimate \citep{Tian2008} makes this \ac{CCO} candidate the only one
that is almost certainly too old for $r$-modes, although we still set upper
limits on $r$-mode amplitude.
The distance estimate is also from \citet{Tian2008}.
We used the first-observation position contained in the name of the candidate
\ac{CCO} rather than the slightly better \textit{Chandra} position reported by
\citet{Halpern2010}; the roughly $1''$ difference is not significant for
\ac{GW} integration times used in this paper.

\textit{G354.4+0.0}---All parameters are from the discovery paper
\citep{Roy2013}.
No associated point source has been detected yet, but if the remnant's age is
correct any young neutron star should be within roughly $20''$ of the center
(whose location we used for the \ac{GW} search).
Such a position error is not significant for the integration times used here.

\textit{Fomalhaut b}---Considered an extrasolar planet candidate since its
discovery in \relax{a} visible light \relax{image} \citep{Kalas2008}, this
has been proposed based on a lack of infrared detection to be a serendipitous
discovery of a nearby neutron star \citep{Fomb}.
Parameters are taken from \citet{Fomb}, with the maximum distance an attempt
to balance the uncertainties in the scenarios discussed there.
After this search was run, \citet{Poppenhaeger2017} searched for and did not
find the object with \textit{Chandra}.
If the object is a neutron star, this somewhat reduced the possible distance
and significantly increased the minimum age.

\subsection{Parameter Space}

After sky position, the key parameters for each search were the \ac{GW}
frequency band $(f_{\min}, f_{\max})$ and time span of integration
$T_\mathrm{span}.$
As in \citet{S6NineSNRs}, these parameters were determined in an iterative
process intended to produce a search more sensitive than $h_0^\mathrm{age}$
over as wide a frequency band as possible for a fixed computational cost.
Due to Doppler shifts and several features of the analysis, we capped the
maximum frequency at 1998~Hz rather than the 2~kHz in the \acp{SFT}.
The cost, approximated as proportional to $a^{-1.1} f_{\max}^{2.2}
T_\mathrm{span}^4,$ was kept comparable to \citet{S5CasA} for most targets,
but Cas~A was allocated ten times as many computational cycles due to its
status as youngest known neutron star in the galaxy.
Due to some inaccuracy in the power-law fit used for computational cost as a
function of the key parameters, the computational cost and sensitivity varied
by up to 20--30\% from these goals.
For a given frequency $f,$ as in \citet{S5CasA} and \citet{S6NineSNRs}, we
searched
\begin{equation}
-\frac{f}{a} \le \dot{f} \le -\frac{1}{6} \frac{f}{a};
\end{equation}
and for a given $\dot{f}$ we searched
\begin{equation}
2\frac{\dot{f}^2}{f} \le \ddot{f} \le 7\frac{\dot{f}^2}{f}.
\end{equation}
These ranges and the computational cost fixed $f_{\min},$ $f_{\max},$ and
$T_\mathrm{span}$ for each search.

We then chose the start time of each search by the same method as
\citet{S5CasA} and \citet{S6NineSNRs}, minimizing the harmonic mean of the
strain noise power spectral density during the span over the frequency band
$(f_{\min}, f_{\max}).$
Neglecting the small effect of the declination of the target, this corresponds
to maximizing the search sensitivity for a fixed $T_\mathrm{span}$---which is
roughly a fixed computational cost.
Hence the algorithm chose spans when both interferometers had good noise
performance and little down time, usually later in \ac{O1}.
The resulting search parameters are described in Table~\ref{t:targets2}.

We applied the same consistency checks as in previous searches:
For each search we checked using the parameter space metric
\citep{Whitbeck2006, Wette2008} that neglect of the third frequency derivative
in Eq.~(\ref{ft}) did not significantly reduce $2\mathcal{F},$ even in the
worst case (G1.9+0.3).
We also checked that the position uncertainties of the targets also did not
significantly reduce $2\mathcal{F}.$
A simple approximation \citep{Whitbeck2006} suggests that the sky resolution
of these searches is an arcminute or two at 2~kHz and a 10~day integration,
and it scales inversely with $f_{\max}$ and $T_\mathrm{span}.$
We spot checked this with injection studies and found it to be accurate.
Given the integration times in Table~\ref{t:targets2}, even the worst position
uncertainty ($20''$ for G354.4+0.0) is well within bounds for a single
directed search.
Finally, we checked that the standard 1800~s \ac{SFT} duration did not
diminish sensitivity to signals with $\dot{f}$ high enough that the frequency
could move to another \ac{SFT} frequency bin over the duration of the
\ac{SFT}.
This effect was negligible except for SNR~G1.9+0.3, where it could reduce the
sensitivity (raise the detectable $h_0$) by of order 10\% at frequencies above
1~kHz.

\subsection{Post-processing}
\label{s:post}

Each search recorded a list of candidates with high values of $2\mathcal{F},$
which was then pared using two automated vetoes designed for instrumental
artifacts, as used in \citet{S5CasA} and \citet{S6NineSNRs}.
The ``Fscan veto'' used a normalized spectrogram formed from the \acp{SFT} to
detect and veto spectral lines and nonstationary noise.
Its implementation and parameters were the same as in \citet{S6NineSNRs}
except that we fixed a bug in the old code whereby the Doppler shift due to
the Earth's orbital motion was not applied.
(This bug allowed more noise lines to pass the automated vetoes and require
manual scrutiny, but had a negligible effect on the false dismissal rate.)
The ``interferometer consistency veto'' ruled out candidates for which a
single-interferometer $2\mathcal{F}$ exceeded the two-interferometer
$2\mathcal{F}$ for the same event, indicating a disturbance present in only
one interferometer.
It also vetoed entire search jobs if the number of candidates vetoed was high
enough.
This veto was also applied in the same way as in \citet{S6NineSNRs}, except
that the threshold for vetoing an entire search job was 5\% of the templates
in that job.
Unlike in previous papers in this series, we also vetoed a list of known
instrumental spectral lines compiled from studies of the interferometers
\citep{Covas2018}.

After these steps, including fixing the Doppler bug, the searches still had
almost two thousand jobs containing non-vetoed outliers above the 95\%
confidence level for Gaussian noise.
All of these jobs were examined by hand.
As in \citet{S6NineSNRs}, two plots were made and inspected for each job.
(See Figure~1 of that paper for illustrative examples.)
In case of a real or injected signal the first plot, of $2\mathcal{F}$ vs.\
frequency for all loud candidates in the job, would show a $\delta$-function
like spike even for very loud signals, as verified by studying hardware
injections.
The candidates generally showed broad bands of high noise, occupying a
fraction of order unity of the search job frequency band except for a handful
which occupied a few percent of the search band.
These few candidates, which were still of order one hundred times broader than
a real signal would be, were verified to be hardware injected test signals
detectable in the wrong sky location due to their huge amplitudes.
The second plot for each search job containing candidates was a semilog
histogram of loud candidates, which on inspection typically showed the tail of
a $\chi^2(4)$ distribution with the wrong amplitude, indicative of a broad
band disturbance in the noise spectrum.
See \citet{S6NineSNRs} for examples and further details.

No candidates survived inspection of these plots, and therefore we conclude
that no astrophysical signal was detected.

\section{Upper limits}

\begin{figure*}
\begin{center}
\includegraphics[width=0.49\textwidth]{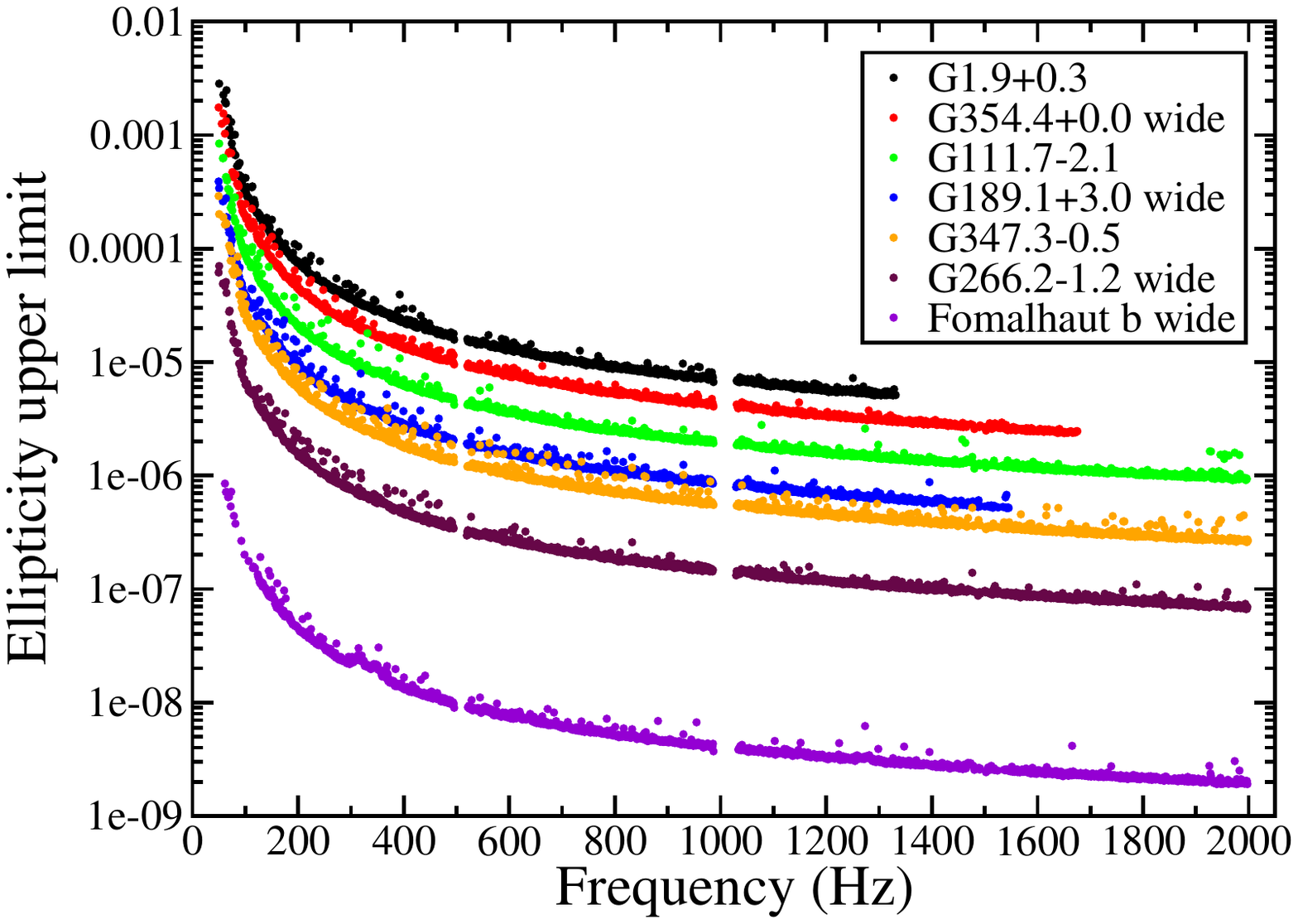}
\includegraphics[width=0.49\textwidth]{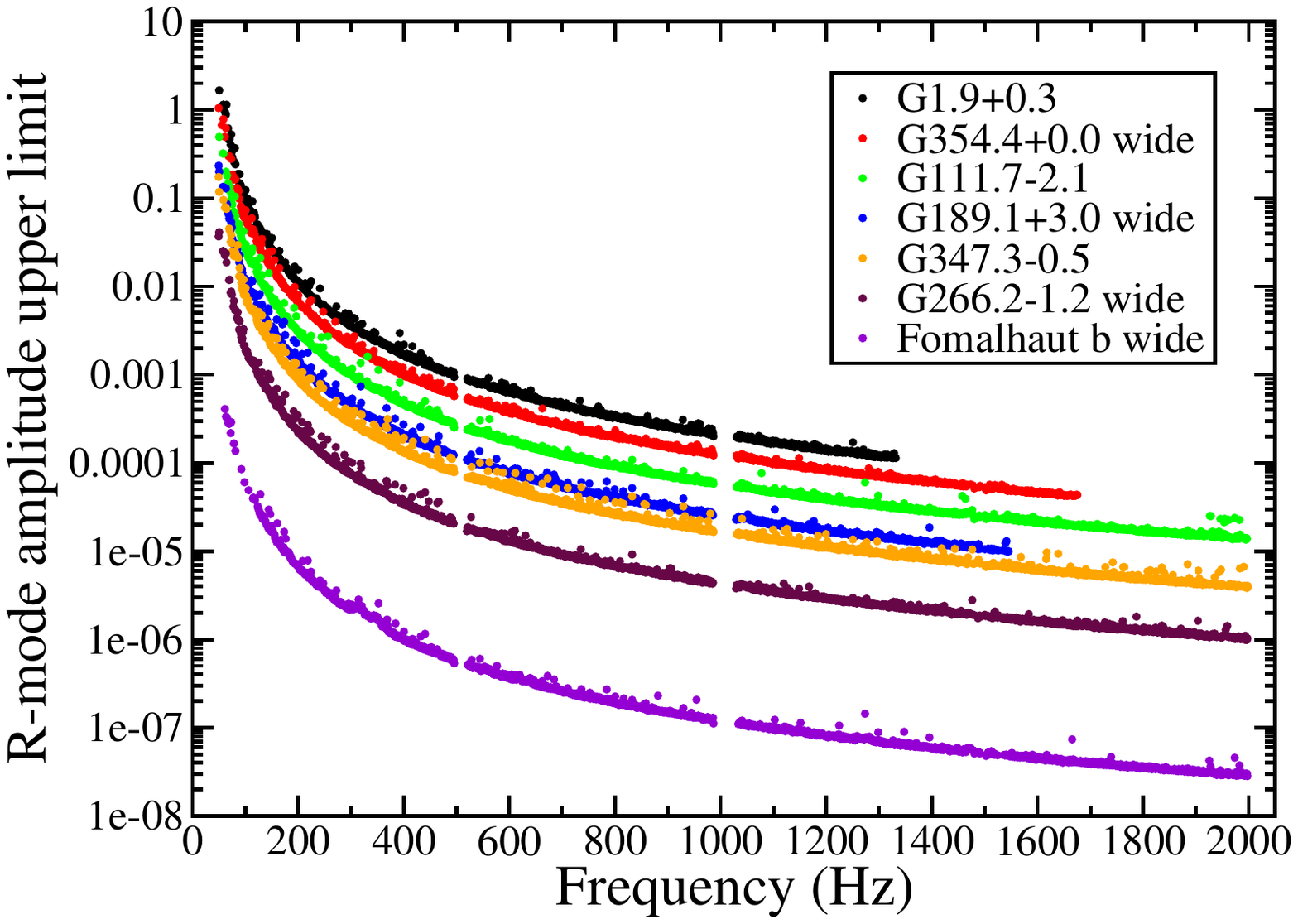}
\end{center}
\caption{
\label{f:eanda}
Upper limits on fiducial ellipticity (left panel) and $r$-mode amplitude
(right panel) for a representative sample of searches.
}
\end{figure*}

Our method of setting upper limits was almost the same as in previous papers
\citep{S5CasA, S6NineSNRs}.
In each 1~Hz band searched, we estimated the value of $h_0$ that would be
detected 95\% of the time by our search (assuming random variation of other
signal parameters such as inclination of the star's rotation axis to the line
of sight) at a louder value than the loudest $2\mathcal{F}$ actually recorded
by the search in that band.
We made an initial estimate from a semianalytic integration of the expected
$2\mathcal{F}$ distribution.
Then we injected simulated signals with different values of $h_0$ near this
value to refine the location of the 95\% confidence (5\% false dismissal)
threshold.
We reduced the number of injections per band to 1000 (from 6000 in previous
papers) due to the computational cost of setting upper limits on wider bands.

For each search we pared the list of upper limits on $h_0$ versus frequency.
We dropped bands where the injections indicated the false dismissal rate was
more than 5\% and we dropped $\pm1$~Hz bands around harmonics of the 60~Hz
power mains up to 300~Hz.

The resulting upper limits on $h_0,$ in 1~Hz frequency bands, are plotted in
Figs.~\ref{f:h01}--\ref{f:h00}.
Each curve has roughly the same shape as the amplitude spectral density of the
strain noise.
The line of dots near the top of each plot corresponds to bands where no upper
limit was set.
Some features such as the ``violin modes'' of the interferometer test mass
suspension (roughly 500~Hz and harmonics) are evident.
The horizontal line in each plot is $h_0^\mathrm{age},$ the strain the search
was intended to beat.
In some cases the estimate of sensitivity made before performing the search
was wrong by of order 10\%, so the upper limits (lower dots) do not always lie
below the line.

Upper limits on $h_0$ can be converted to upper limits on fiducial
neutron-star ellipticity $\epsilon = |I_{xx} - I_{yy}| /I_{zz}$ (where
$I_{ab}$ is the moment of inertia) using \citep[e.g.][]{Wette2008}
\begin{equation}
\epsilon = 9.5\times10^{-5} \left( \frac{h_0} {10^{-24}} \right)
\left( \frac{D} {\mbox{1 kpc}} \right) \left( \frac{\mbox{100 Hz}} {f}
\right)^2.
\end{equation}
This number assumes $I_{zz} = 10^{45}$\,g\,cm$^2.$
Uncertainties in the mass, radius, and neutron star equation of state make the
conversion from $h_0$ to $\epsilon$ uncertain by a factor of two or more.
This fiducial ellipticity can be converted to the true shape of the star
\citep{JohnsonMcDaniel2013} or other quantities \citep{Owen2010}.
We plot upper limits on $\epsilon$ for a selection of searches in the left
hand panel of Fig.~\ref{f:eanda}.
We do not plot the indirect limits on $\epsilon$ and $\alpha$ derived from
$h_0^\mathrm{age}$ since they are close to the direct upper limits on the
scale of the plot.
We do not plot the remaining searches because their upper limits are close to
those of the searches plotted.
The great differences between curves are mainly due to the distances to the
sources; hence Fomalhaut~b has the best upper limits---of order $10^{-9}$ at
high frequencies.

Upper limits on $h_0$ can be converted to the common $r$-mode amplitude
parameter $\alpha$ \citep{Lindblom1998} via \citep{Owen2010}
\begin{equation}
\alpha = 0.028 \left( \frac{h_0} {10^{-24}} \right) \left( \frac{\mbox{100 Hz}}
{f} \right)^3 \left( \frac{D} {\mbox{1 kpc}} \right).
\end{equation}
This number assumes a fiducial set of stellar parameters described in
\citet{Owen2010} and is uncertain by a factor of up to about three depending
in the neutron star mass and equation of state.
We plot upper limits on $\alpha$ for a selection of searches in the right hand
panel of Fig.~\ref{f:eanda}.
Again, the differences between curves are mainly due to the source distances.
The best upper limits, apart from Fomalhaut~b which is almost certainly too
old for active $r$-modes, are of order $10^{-6}$ at high frequencies for
Vela~Jr.

\section{Discussion}

These are the first directed searches of Advanced LIGO data using continuous
wave analysis methods.
These searches have improved on previous directed searches by covering wider
parameter ranges and more targets, and by setting better upper limits on
targets searched previously.
Our upper limits on $h_0$ approach $2\times10^{-25}$ for many targets and
approach $1\times10^{-25}$ for one---about a factor of 3 improvement on
\citet{S6NineSNRs}, due mainly to the improvement in the detectors.
And our upper limits beat the indirect limit $h_0^\mathrm{age}$ over bands of
1--2~kHz for more targets than were ever published before.
[Searches for some of these targets in less sensitive S6 data for the purpose
of testing code were described in an unpublished thesis \citep{Idrisy}.]
As with previous data runs, we improved on the sensitivity of all-sky
wide-band searches \citep{O1FullBandAllSky} but did not match the sensitivity
of searches for known pulsars with full timing solutions \citep{O1Pulsars}.
As before, the directed searches described here also have the caveats that
there might be no neutron star present in some cases, any neutron star might
be spinning too slowly to be detected, and a neutron star spinning at a
detectable frequency might glitch---the latter phenomenon in a \ac{CCO} is now
an observation \citep{Gotthelf2018} rather than a surmise, and would somewhat
reduce the sensitivity of these searches \citep{Ashton2017}.
\relax{Even with some longer integration times here, timing noise is not an
issue unless these objects are orders of magnitude noisier than known pulsars
\citep{Ashton2015}.}

Most of our upper limits on $\epsilon$ and $\alpha$ are competitive with the
largest numbers predicted by theory.
The maximum $\epsilon$ for ``mountains'' supported by elastic stresses of
normal neutron star matter is probably $10^{-5}$--$10^{-6}$
\citep{Horowitz2009, JohnsonMcDaniel2013a, Baiko2018}, and for many of our
searches upper limits are in this region over hundreds of~Hz.
The maximum $\alpha$ (nonlinear saturation amplitude) for $r$-modes is
probably of order $10^{-3}$ \citep{Bondarescu2009}, and for many of our
searches upper limits beat this over hundreds of~Hz.
Mountains supported by an internal magnetic field can produce $\epsilon$ of
order $10^{-4} (B/10^{15}\mbox{ G})^2$ where $B$ is the poloidal part of the
field \citep[e.g.][]{Ciolfi2013}.
Since, unlike elastic mountains, magnetic mountains are likely to be within
about an order of magnitude of this limit for a given internal field,
depending on its configuration, our upper limits on $\epsilon$ translate into
rough limits on internal magnetic field---if a neutron star is present and
spinning rapidly enough to emit \acp{GW} in band.

More data from Advanced LIGO and Advanced Virgo is now available, with more
live time and lower noise amplitude than before.
\relax{The detectable values of intrinsic strain, ellipticity, and $r$-mode
amplitude are proportional to the noise amplitude and to the inverse square
root of the live time.}
This makes more targets feasible for directed searches at greater sensitivity,
increasing the chances of a detection of continuous \acp{GW}.
Such searches will be done in the near future.

The authors gratefully acknowledge the support of the United States
National Science Foundation (NSF) for the construction and operation of the
LIGO Laboratory and Advanced LIGO as well as the Science and Technology Facilities Council (STFC) of the
United Kingdom, the Max-Planck-Society (MPS), and the State of
Niedersachsen/Germany for support of the construction of Advanced LIGO 
and construction and operation of the GEO600 detector. 
Additional support for Advanced LIGO was provided by the Australian Research Council.
The authors gratefully acknowledge the Italian Istituto Nazionale di Fisica Nucleare (INFN),  
the French Centre National de la Recherche Scientifique (CNRS) and
the Foundation for Fundamental Research on Matter supported by the Netherlands Organisation for Scientific Research, 
for the construction and operation of the Virgo detector
and the creation and support  of the EGO consortium. 
The authors also gratefully acknowledge research support from these agencies as well as by 
the Council of Scientific and Industrial Research of India, 
the Department of Science and Technology, India,
the Science \& Engineering Research Board (SERB), India,
the Ministry of Human Resource Development, India,
the Spanish  Agencia Estatal de Investigaci\'on,
the Vicepresid\`encia i Conselleria d'Innovaci\'o, Recerca i Turisme and the Conselleria d'Educaci\'o i Universitat del Govern de les Illes Balears,
the Conselleria d'Educaci\'o, Investigaci\'o, Cultura i Esport de la Generalitat Valenciana,
the National Science Centre of Poland,
the Swiss National Science Foundation (SNSF),
the Russian Foundation for Basic Research, 
the Russian Science Foundation,
the European Commission,
the European Regional Development Funds (ERDF),
the Royal Society, 
the Scottish Funding Council, 
the Scottish Universities Physics Alliance, 
the Hungarian Scientific Research Fund (OTKA),
the Lyon Institute of Origins (LIO),
the Paris \^{I}le-de-France Region, 
the National Research, Development and Innovation Office Hungary (NKFIH), 
the National Research Foundation of Korea,
Industry Canada and the Province of Ontario through the Ministry of Economic Development and Innovation, 
the Natural Science and Engineering Research Council Canada,
the Canadian Institute for Advanced Research,
the Brazilian Ministry of Science, Technology, Innovations, and Communications,
the International Center for Theoretical Physics South American Institute for Fundamental Research (ICTP-SAIFR), 
the Research Grants Council of Hong Kong,
the National Natural Science Foundation of China (NSFC),
the Leverhulme Trust, 
the Research Corporation, 
the Ministry of Science and Technology (MOST), Taiwan
and
the Kavli Foundation.
The authors gratefully acknowledge the support of the NSF, STFC, MPS, INFN, CNRS and the
State of Niedersachsen/Germany for provision of computational resources.

This paper has been assigned document number LIGO-P1800333-v7.
\relax{Data files for the figures are available
at~\texttt{https://dcc.ligo.org/P1800333/public}.}

\relax{\software{LALSuite \citep{lalsuite}}}

\bibliography{paper}

\end{document}